# An infection-responsive collagen-based wet-spun textile fibre for wound monitoring

Jonathon Gorman,[1,2] Charles Brooker,[1,2] Xinyu Li,[1] Giuseppe Tronci[1,2*]

[1] Clothworkers Centre for Textile Materials Innovation for Healthcare (CCTMIH), Leeds Institute of Textiles and Colour (LITAC), University of Leeds, United Kingdom

[2] Division of Oral Biology, St. James's University Hospital, School of Dentistry, University of Leeds, United Kingdom

## Abstract

Wound infections are a significant clinical and socioeconomic challenge, contributing to delayed healing and increased wound chronicity. To enable early infection detection and inform therapeutic decisions, this study investigated the design of pH-responsive collagen fibres using a scalable wet spinning process, evaluating product suitability for textile dressings and resorbable sutures. Type I collagen was chemically functionalised with 4-vinylbenzyl chloride, enabling UV-induced crosslinking and yielding mechanically robust fibres. Bromothymol blue, a halochromic dye responsive to pH changes, was incorporated via drop-casting to impart visual infection-responsive colour change. Gravimetric analysis and Fourier Transform Infrared Spectroscopy confirmed high dye loading, whereby a Loading Efficiency of 99±3 wt.% was achieved. The fibres exhibited controlled swelling in aqueous environments (Swelling Ratio: 323±79—492±73 wt.%) and remarkable wet-state Ultimate Tensile Strength (UTS: 12±3—15±7 MPa), while up to ca. 30 wt.% of their initial crosslinked mass was retained after 24 hours in a collagenase-rich buffer (pH 7.4, 37°C, 2 CDU) and ethanol series dehydration. Importantly, distinct and reversible colour transitions were observed between acidic (pH 5) and alkaline (pH 8) environments, with up to 88 wt.% dye retention following 72-hour incubation. The fibres were successfully processed into woven dressing prototypes and demonstrated knotting ability suitable for suture applications. Overall, these wet-spun collagen fibres integrate infection-responsive capability, biodegradability, and scalable fabrication, representing a promising platform for smart wound dressings and resorbable sutures.

## Graphical abstract

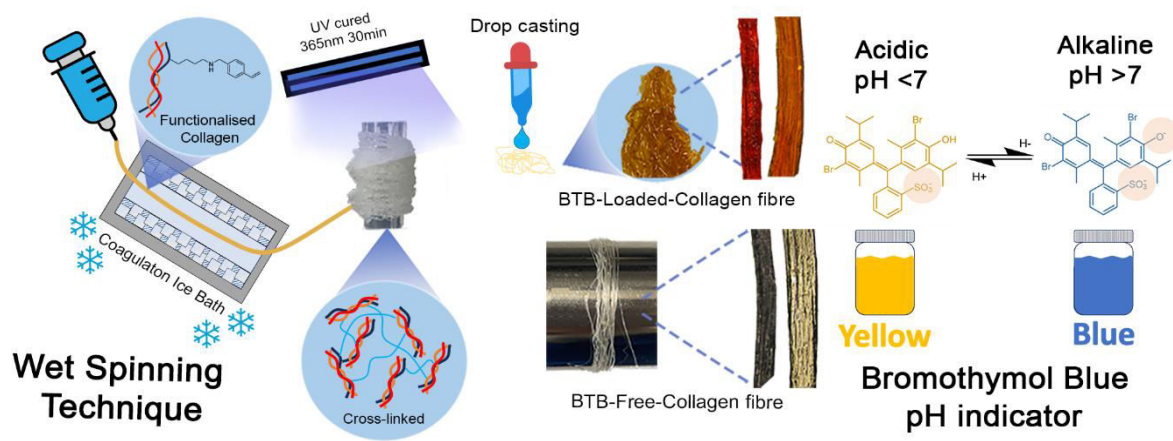

---

[*] *Address*: Level 7 Wellcome Trust Brenner Building, St. James's University Hospital, Leeds, LS9 7TF. *Email correspondence*: g.tronci@leeds.ac.uk

**Key Words:** Collagen fibre, Wet spinning, Infection monitoring, Wound dressing, Bromothymol blue, pH indicator

**Highlights**

- Individual collagen-based textile fibres are generated via scalable wet spinning

- Fibres reveal controlled swelling and remarkable wet-state tensile properties

- Fibre drop casting with bromothymol blue enables infection-responsive colour change

- Fibres can be woven into a dressing prototype and show knot-forming/tying abilities

## 1. Introduction

Wound infections remain a major clinical challenge, contributing significantly to delayed healing, prolonged hospital visits, and increased healthcare costs worldwide. In severe cases, infected wounds can lead to sepsis and require limb amputation [1]. The cost of treating wounds in the UK alone was estimated to be £8.8 billion in 2022 [2], and in the US, this figure approaches $150 billion [3]. Conventional methods for diagnosing wound infections, such as microbial swabbing, tissue biopsies, and visual inspection, are often invasive, slow to yield results, expensive, or subjective [4]. These shortcomings have prompted an increased interest in non-invasive, real-time biomarkers that can signal infection progression at earlier stages.

One promising biomarker of wound infection is local pH, which undergoes characteristic changes during wound progression. Monitoring wound pH is a key indicator for predicting bacterial infections and assessing the stages of wound healing. Healthy skin typically exhibits an acidic pH (≈4–6) [5], which is associated with the production of lactic acid and this acidic pH is beneficial for enhancing fibroblast proliferation, angiogenesis, and collagen synthesis towards wound healing [6]. On the other hand, infected wounds become increasingly alkaline due to the destruction of the extracellular matrix (ECM) and the release of ammonia, leading to an increased wound alkalinity (pH > 7) [7]. Consequently, several wound monitoring technologies have emerged to address this need, including electrochemical sensors [8], colorimetric assays [9], and optical probes [10].

While promising, many of these 'smart' wound monitoring systems still need to overcome technical challenges. For example, conventional glass-based electrochemical sensors used to monitor pH are often mechanically fragile [11], and fluorometric methods often lack long-term stability and require the use of a UV lamp [12, 13]. Ion-sensitive field-effect transistors (ISFETs) represent another class of electrochemical pH sensors [14]; however, concerns remain regarding their relatively high operating voltage and limited sensitivity [15]. The widespread clinical adoption of these smart wound dressings remains limited, largely due to factors such as high cost, lack of disposability, and bulkiness, which hinder their seamless integration into conventional dressings [16]. Moreover, the design complexity presents additional barriers to clinical translation, including regulatory approval and user compliance [17], highlighting the need for simple, robust designs that are both scalable and clinically viable.

Type I collagen is the most abundant protein of connective tissues in the human body and widely recognised for its biocompatibility, biodegradability, and low immunogenicity. These characteristics make it one of the most deployed biomaterials in commercially available wound care products, including wound dressings and surgical sutures, due to its wound healing and inherent bioactive properties. Type I collagen helps to simulate the native microenvironment of a wound and aids in promoting cell proliferation, differentiation, and migration, which is essential for stimulating the synthesis of new ECM and attracting fibroblasts to the wound bed [18-20]. Building on the advantages of textile-based wound care platforms, including wound dressings and sutures, type I collagen fibres present an ideal biodegradable substrate to support wound healing.

Unlike purely synthetic textiles, type I collagen degrades enzymatically (via collagenases), releasing chemotactic peptides that can promote angiogenesis and tissue remodelling [21-23]. Additionally, the presence of reactive side-chains allows for chemical functionalisation [24], potentially enabling the integration of structural support, real-time sensing, and/or controlled therapeutic release within a single, biocompatible platform. The limited solubility of type I collagen in organic solvents also makes it compliant with scalable fibre manufacturing techniques, i.e. wet spinning, aiming to accomplish individual fibres as building block of fibrous materials. On the other hand, the uncontrollable water-induced swelling of type I collagen *ex vivo* present significant challenges towards the successful delivery of collagen-based materials at industrial scale. To overcome these issues, there is a crucial need to develop multiscale design approaches that enable control of molecular scale and microscale in collagen-based materials aiming to meet key functional requirements and usability needs.

Many commercial wound dressings are made from textile fibres due to their inherent flexibility and mechanical strength [25-27]. By embedding responsive elements directly into fibres, smart textiles can provide real-time feedback on wound status while maintaining breathability and a close, non-disruptive interface with the wound bed [28]. In addition to dressings, textile fibres can also be integrated into sutures, which are in direct and prolonged contact with tissue, offering a valuable platform for real-time monitoring, targeted therapeutic delivery alongside biodegradability, removing the requirement for secondary surgery [29-32]. Utilising scalable manufacturing techniques, such as wet spinning [33] and weaving [34], can support the production of disposable, low-cost devices suitable for clinical translation [35-37].

In our previous work, we developed a pH-responsive collagen-based theranostic dressing made of UV-cured functionalised type I collagen, incorporating bromothymol blue (BTB). The resulting prototype demonstrated cellular tolerability, dye retention over clinically relevant timeframes, and prompt visual infection indication [38]. However, these formats were limited in their integrability with use-inspired clinical functionalities. In this work, we focus on the design of the functionalised collagen product in fibrous form, which offers several practical and translational advantages. To achieve this, we identified wet spinning as a promising, scalable manufacturing route to produce continuous collagen fibres with controlled fibre morphology, thermo-mechanical properties, and UV-cured molecular network architecture, so that their applicability as three-dimensional textiles and resorbable suture was successfully demonstrated. Wet spun functionalised collagen fibres were loaded with BTB via a gravitational drop cast method, allowing the fibres to endow pH-sensitive characteristics for use in textile wound dressings and resorbable sutures. The molecular configuration of BTB enables visual colour changes in response to variations in environmental pH without the need for clinical equipment or an algorithm. This multiscale design approach aims to overcome limitations observed in collagen-based wound care systems, such as poor fibre integrity in aqueous environments, rapid dye release, and limited handling capability, while supporting the development of smart wound monitoring devices aligned with industrial manufacturing practice and regulatory expectations.

## 2. Materials and methods

### 2.1 Materials

Rat tails were provided post-mortem from the School of Dentistry, University of Leeds and used for the extraction of type I collagen via acidic treatment, as described by Rajan et al. [39]. Triethylamine (TEA), ethanol (EtOH), picryl sulfonic acid solution (5% w/v in water), sodium bicarbonate, calcium chloride, and collagenase from *Clostridium histolyticum* (125 CDU·mg$^{-1}$) were purchased from Sigma-Aldrich (Gillingham, UK). A 17.4M acetic acid (AcOH) solution, and glacial hydrochloric acid (HCl) solution, were also purchased from Sigma-Aldrich and diluted with deionised water before use. 4-vinylbenzyl chloride (4VBC) was purchased from Thermo Fisher Scientific (Massachusetts, USA). 2-Hydroxy-4′-(2-hydroxyethoxy)-2-methylpropiophenone (I2959) and N-[Tris(hydroxymethyl)methyl]-2-aminoethanesulfonic acid (TES) were purchased from Merck (Feltham, UK). Phosphate-buffered saline (PBS) was purchased from Corning UK (Flintshire, UK). Polysorbate 20 (Tween 20) was purchased from Scientific Laboratory Supplies Ltd (Nottingham, UK). Bromothymol Blue (BTB) was purchased from Alfa Aesar (Heysham, UK).

### 2.2 Manufacture of collagen fibres

4VBC-functionalised collagen was synthesised as previously reported by Tronci et al. [40]. Briefly, rat tail collagen (CRT) was dissolved in 17.4 mM AcOH and reacted with 25-molar excess of 4VBC and TEA with respect to the molar content of collagen lysines (≈ 2.7×10$^{-4}$ mol·g$^{-1}$). After 24 hours, the reaction mixture was precipitated in 10-fold ethanol for 24 hours, recovered by centrifugation, and air dried. The resulting 4VBC-functionalised collagen product was dissolved in 17.4 mM AcOH supplemented with 1 wt.% I2959 under magnetic stirring for 24 hours. The resulting wet-spinning dope was loaded into a 10 mL syringe with an internal diameter of 15.7 mm and equipped with a 1.1 mm (19G) diameter needle. The solution was thoroughly agitated to minimise the risk of air bubbles and mounted onto a syringe pump (World Precision Instruments, Hitchin, UK). Wet spinning was performed at a flow rate of 12.5 mL·hr$^{-1}$ by immersing the needle into a coagulation bath comprising either EtOH or EtOH supplemented with 0.5-1 wt.% I2959, maintained in contact with ice. A thin layer of petroleum jelly was applied to the bottom of the coagulation bath to prevent fibre adhesion. The extruded fibres were left in the coagulation bath for 5-10 minutes (until they became transparent), followed by 15 minutes of UV irradiation (1.8 mW·cm$^{-1}$, Chromato-Vue C-71, Analytik Jena, Upland, CA, USA), on both sides. Samples were then dehydrated via a graded distilled water-ethanol series (20, 40, 60, 80, 100 vol.% EtOH) prior to air-drying. UV-cured wet-spun collagen fibres were designated as F-4VBC*, where *F* indicates the fibre format, *4VBC* denotes chemical functionalisation, and * indicates the UV-cured, crosslinked state.

### 2.3 Fibre loading with BTB

A gravimetric method was employed to quantify the loading content of BTB in drop-cast collagen fibres (n = 14). The initial mass of each individual dry sample ($m_i$) was recorded using a precision five-decimal-place analytical balance (Sartorius, Göttingen, Germany). Subsequently, 100 μL of a 0.2 wt.% BTB solution was added to each sample using a micropipette. After 48 hours of air-drying, the final mass ($m_f$) of the drop-cast samples was recorded, and the Loading Efficiency (LE) calculated via Eq. (1):

$$LE = \frac{(m_f - m_i)}{m_{BTB}} \qquad \textit{Equation 1}$$

where $m_{BTB}$ is the mass of BTB contained in the volume of the aqueous solution applied to the samples during drop casting. Drop-cast, BTB-loaded fibres were coded as B-4VBC*, where *B* refers to the fibre encapsulation with BTB, while *4VBC* and *\** have the same meaning as above.

**2.4 Chemical characterisation**

Attenuated total reflectance Fourier-transform infrared (ATR-FTIR) spectroscopy (Perkin Elmer Spectrum 3) was employed to assess the chemical composition of the dry samples (n = 3). The FTIR spectra were recorded with a spectral resolution of 4 cm$^{-1}$ and a scan interval of 2 cm$^{-1}$.

2,4,6-Trinitrobenzenesulfonic acid (TNBS) assays were conducted as a colorimetric technique to quantify primary amino groups in proteins [41]. Solutions of 0.5 wt.% TNBS and 4 wt.% NaHCO$_3$ were prepared in distilled water. One millilitre of each solution was pipetted onto samples of native and functionalised collagen (11 mg, n = 3). After 4 hours of stirring at 40°C, 3 mL of 6N HCl was added, followed by 1 hour of incubation at 60°C. Following equilibration to room temperature, 5 mL of distilled water was added to all samples. Samples were extracted three times with 20 mL of diethyl ether. Subsequently, 5 mL of the aqueous phase were collected and diluted with 15 mL of distilled water prior to absorbance measurement at 346 nm using a UV-Vis spectrophotometer (Jenway 6305 UV-vis spectrophotometer, Essex, United Kingdom). Blank samples (11 mg) were prepared in the same manner as the test samples, except that 3 mL of 6N HCl was added together with the TNBS and NaHCO$_3$ solutions prior to incubation at 40°C. The molar content of collagen lysines and the degree of collagen functionalisation (F) were quantified using Eq. (2) and (3):

$$\frac{Mol(lys)}{g(collagen)} = \frac{2 \times ABS \times V}{\mu \times b \times m_d} \qquad \text{Equation 2}$$

$$F = \left(1 - \frac{mol(Lys)_{Funct}}{mol(Lys)_{Native}}\right) \times 100 \qquad \text{Equation 3}$$

whereby ABS is the solution absorbance recorded at 346 nm, μ is the molar absorption coefficient of 2,4,6-trinitrophenyl lysine (1.46×10$^4$ M$^{-1}$·cm$^{-1}$), b is the path length of the cuvette (1 cm), m$_d$ is the mass of the dry sample (11 mg) and V is the volume of solution (0.02 litres).

**2.5 Thermal analysis**

Thermogravimetric analysis (TGA) and differential scanning calorimetry (DSC) were performed to assess the thermal properties of dry BTB-free and BTB-loaded collagen fibres. TGA (Perkin Elmer TGA 4000, Massachusetts, USA) was carried out under a nitrogen flow rate of 50 cm³·min$^{-1}$, starting at 30°C and ramping to 900°C at 10°C·min$^{-1}$. Samples were held at 900°C for 5 minutes in a nitrogen atmosphere, followed by 1 minute in air at the same temperature.

DSC (Perkin Elmer DSC 4000, Massachusetts, USA) was performed on dehydrated samples (n = 3, ≈10 mg). Thermograms were recorded from -10°C to 150°C at a heating rate of 10°C·min$^{-1}$. Prior to measurement, the DSC cell was calibrated using indium as the standard under a nitrogen flow rate of 50 cm³·min$^{-1}$.

**2.6 Optical and electron microscopy**

Light microscopy (Leica M205 C) was used to measure the thickness of dehydrated and rehydrated fibres (n = 10)*,* with rehydration performed in PBS for 10 hours. Fibre morphology was examined in the dry state using scanning electron microscopy (SEM; Hitachi SU3900, Hitachi, Tokyo, Japan) under high vacuum mode. Prior to imaging, all samples were gold-coated using a sputter coater (Emscope SC500, Emscope, Ashford, United Kingdom).

### 2.7 Quantification of linear density

The dry mass ($m_d$) of the fibre samples was measured (Ohaus Discovery DV314C, New Jersey, USA), while their length (l) was measured with a universal tensile machine (Instron 5544-Massachusetts, USA). The linear density of the fibres ($\lambda$, tex) was then calculated using Eq. (4):

$$\lambda = \frac{m_d}{l} \times 1000 \qquad \text{Equation 4}$$

### 2.8 Swelling ratio and gel content measurements

The dry mass ($m_d$) of BTB-free and BTB-loaded fibres was recorded using an analytical balance (Ohaus Discovery DV314C, New Jersey, USA). Samples were then incubated in deionised water (DI) for 2 hours. After incubation, swollen samples were gently blotted on filter paper, and the swollen mass ($m_s$) was recorded. The swelling ratio (SR) was calculated using the Eq. (5):

$$SR = \frac{m_s - m_d}{m_d} \times 100 \qquad \text{Equation 5}$$

In addition, the gel content (G) of the fibres (n = 3) was quantified as an indirect measurement of the crosslink density of the respective UV-cured molecular network. Dry samples of known mass ($m_d$) were incubated in 17.4 mM AcOH for 24 hours. Following incubation, the samples were collected and dried through a graded ethanol-distilled water series before measuring the resulting mass ($m_e$). G was then calculated using Eq. (6):

$$G = \frac{m_e}{m_d} \times 100 \qquad \text{Equation 6}$$

### 2.9 Tensile testing of hydrated fibres

The tensile properties of the individual BTB-free and BTB-loaded collagen fibres were assessed after incubation in PBS for 2 hours, using an Instron 5544 (Massachusetts, USA) at a temperature of 20 ± 2°C and a relative humidity of 65 ± 4% (n = 5). The fibres had a gauge length of 20 mm, the crosshead speed was 2 mm·min$^{-1}$, and tests were conducted using a 5 N load cell. The Young's modulus was calculated using the linear region of the stress-strain curve.

### 2.10 *In vitro* studies of dye release

The dye retention in the drop-cast collagen fibres was indirectly assessed by measuring the dye release following sample incubation in aqueous environments. McIlvaine solutions adjusted to either pH 5 or pH 8 were selected to simulate active and infected wound environments, respectively [5]. Individual collagen fibres (n = 3), loaded with either 100 μg or 60 μg of BTB, were incubated in 5 mL of the aforementioned acidic and alkaline McIlvaine solutions, respectively, at room temperature. The release of BTB in each solution was determined over 4 days by measuring the absorbance of the buffer solution via UV-Vis spectrophotometry (Jenway 6305 UV-vis spectrophotometer, Essex, United Kingdom). Calibration curves were built by dissolving known amounts of BTB in a McIlvaine solution adjusted to either pH 5 or pH 8 (Figure S1, Supp. Inf.), with absorbance recorded at 432 nm and 616 nm, respectively. The resulting calibration curves were used to quantify the BTB released over time.

### 2.11 Colorimetric analysis

The colour of the fibres was recorded using an SF600 Plus-CT spectrophotometer (Datacolor, Lucerne, Switzerland), following UV curing, drop casting, or 96-hour incubation in either pH 5 or pH 8 McIlvaine solution. After calibration against standard black and white standards, the lightness, chroma, and hue were measured using a 3 mm aperture, taking three individual measurements per sample (n = 3). Results were reported as mean ± standard deviation.

### 2.12 Enzymatic degradation tests

Collagenase from *Clostridium histolyticum* (125 CDU·mg$^{-1}$) was employed to assess the enzymatic degradation of BTB-loaded collagen fibres *in vitro*, in accordance with previous reports [31, 42]. Bovine type I collagen was used to fabricate UV-cured wet-spun fibres, due to its high purity and comparable chemical composition to rat tail collagen [43] (Figure S2, Supp. Inf.). Dry BTB-loaded fibres (m$_d$: 10–20 mg; BTB: 60 µg) were incubated in a 1 mL of collagenase-rich buffer (50 mM TES, 0.36 mM calcium chloride, pH 7.4; 2 CDU·mL$^{-1}$) under mild agitation (150 rpm, 37˚C). The value of collagenase concentration was calculated according to Equations S1 and S2 (Supp. Inf.). Enzymatic degradation studies were also performed on native rat tail collagen, UV-cured rat tail collagen fibres, and UV-cured rat tail collagen thin films, following incubation in 5 mL of a collagenase-rich buffer (50 mM TES, 0.36 mM calcium chloride, pH 7.4; 0.4 CDU·mL$^{-1}$) under mild agitation (150 rpm, 37˚C). At selected time points, samples were collected, washed in distilled water and dehydrated through a graded distilled water-ethanol series (0, 20, 40, 60, 80, 100 vol.% EtOH) before air-drying. The dry mass of the retrieved samples (m$_t$) was recorded, and the relative mass (µ$_{rel}$) calculated using Eq. (7):

$$\mu_{rel} = \frac{m_t}{m_d} \qquad \text{Equation 7}$$

where m$_t$ represents the mass of the dry sample at the incubation time t; m$_d$ represents the samples mass prior to incubation.

### 2.13 Fibre weaving and suturing trials

To demonstrate the fibre usability in textile wound dressings, three UV-cured fibres were twisted into yarns, and the yarns spliced to reach a suitable length for further processing. Hand weaving was subsequently carried out to generate a textile wound dressing prototype.
For suture application, individual UV-cured fibres were secured to suture needles (3-0 20 mm 1/2c) to produce single knots, alongside a commercially available silk-braided, non-absorbable suture control.

### 2.14 Statistical analysis

OriginPro 2024b was used to perform statistical analysis and plotting of the experimental data visualisation. Normality was assessed by the Shapiro-Wilk test, followed by one-way ANOVA to evaluate statistical significance (p < 0.05). Data are expressed as mean ± standard deviation.

## 3. Results and discussion

### 3.1 Chemical characterisation

The degree of functionalisation in 4VBC-reacted collagen samples was quantified via a TNBS colorimetric assay and found to be 33 ± 3 mol.% (Table S1, Supp. Inf.), in line with previous reports [31]. This result indicates successful covalent attachment of 4VBC residues onto the lysine side chains and amino termini of collagen. This modification constitutes a critical step towards the formation of a covalently crosslinked network of collagen molecules upon subsequent UV irradiation.

The results of the TNBS assay were complemented by FTIR analysis to characterise the chemical composition of the wet-spun collagen fibres, following either UV exposure or BTB loading. Distinct amide bands were identified in the FTIR spectrum of native type I collagen control (Figure 1A), including: (i) amide A and B at 3300 and 3080 cm$^{-1}$, respectively, indicative of N–H stretching vibrations; (ii) amide I and II bands, at 1630 and 1540 cm$^{-1}$, corresponding to C=O stretching vibrations, and N–H bending and C–N stretching vibrations, respectively; and (iii) an amide III band centred at 1240 cm$^{-1}$, attributed to the C–N stretching and N–H bending vibrations of amide linkages, as well as wagging vibrations of CH$_2$ groups in the glycine backbone and proline side chains. These characteristic signals were also observed in the FTIR spectra of the wet-spun fibres (Figure 1B-C), confirming their collagen-based chemical composition, whereby the detection of the amide III band at 1240 cm$^{-1}$ supports the presence of the triple helical configuration of type I collagen [44].

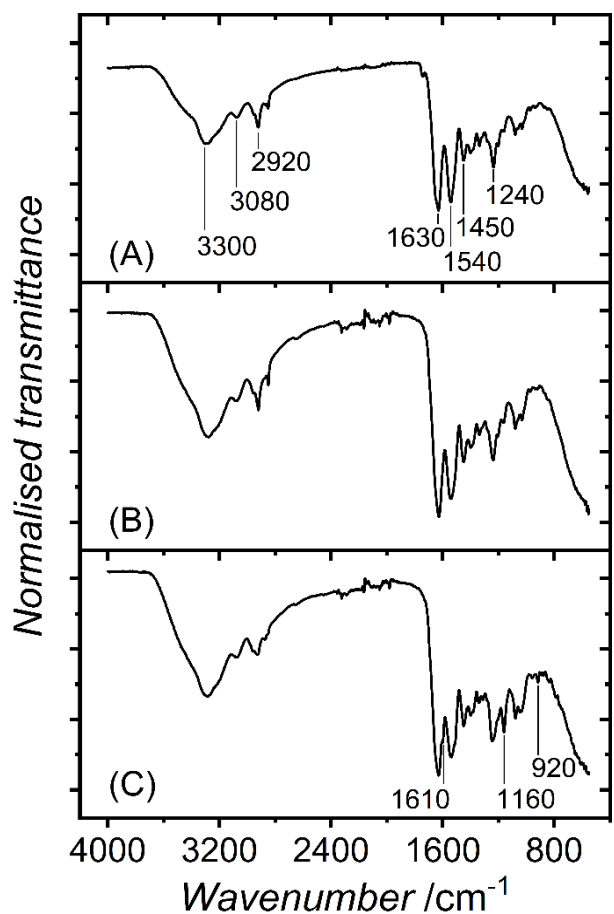

**Figure 1.** FTIR spectra of native type I collagen (A), BTB-free collagen fibre F-4VBC* (B), and BTB-loaded fibre B-4VBC* (C).

The absorbance ratio of the amide III band to the one at 1450 cm$^{-1}$ ($A_{III}/A_{1450}$) was subsequently measured across the collagen control and wet-spun samples to provide further insight into the protein structure within the fibres. A value close to unity was observed in native collagen and UV-cured wet-spun fibres ($A_{III}/A_{1450}$= 0.97–1.02), indicating preservation of the triple helix configuration, consistent with the minimal impact of the selected wet spinning and UV curing process on the collagen molecules [40, 45, 46]. In contrast, BTB-loaded collagen samples exhibited a slightly lower absorbance ratio ($A_{III}/A_{1450}$= 0.89), suggesting partial loss of the collagen triple helices following BTB drop casting. Similar observations have previously been recorded with cast films of the same collagen composition [38] and may be attributed to functional groups in the BTB molecule potentially interfering with the secondary interactions that stabilise the collagen triple helix. This hypothesis is supported by the appearance of an additional shoulder at 1610 cm$^{-1}$ in the FTIR spectra of B-4VBC* compared to F-4VBC*, indicative of secondary interactions between the aromatic rings of 4VBC-functionalised collagen molecules and BTB in the dry state.

In addition to the collagen signals, further bands were detected in the FTIR spectrum of the BTB-loaded sample (Figure 1C), specifically at 1160 and 920 cm$^{-1}$, which were assigned to C–O stretching vibrations and aromatic C-H bending vibrations in BTB, respectively [47]. These observations are supported by gravimetric analysis performed on the same samples, which revealed a weight increase and a BTB loading efficiency of 99 ± 3 wt.% following drop casting (n = 14; Table S2, Supp. Inf.). These results therefore confirm the validity of the selected drop-casting method as a compelling, low-cost and effective technique to accomplish BTB-loaded samples of wet-spun fibre with a high loading efficiency. While we have previously demonstrated the preparation of BTB-loaded cast films (∅: 10-15 mm) [38], these results provide further evidence that this BTB encapsulation strategy is also effective with samples of textile fibres with significantly decreased diameter (∅ << 1 mm).

**3.2 Wet spinning of photoactive collagen fibres**

Wet spinning of 4VBC-functionalised collagen was performed by dissolving the product in a 17.4 mM AcOH solution supplemented with I2959 (1 wt.%), with the aim of producing photoinitiator-containing wet-spun fibres suitable for subsequent UV curing. Given the high solubility of I2959 in ethanol, varying concentrations of the photoinitiator were introduced to the ethanol-based coagulation bath to minimise the risks of photoinitiator diffusion away of the fibre-forming collagen jet. The photoinitiator retention in the fibres was considered crucial due to the direct influence of I2959 concentration on crosslink density, gel content, and mechanical properties of photoinduced hydrogel networks [31, 48, 49]. UV-Vis measurements were therefore conducted on the coagulation bath before and after collagen dope extrusion to quantify any soluble factor release occurring following the wet spinning process (Table S3, Supp. Inf.).

Control experiments performed with I2959-free collagen dope and I2959-free coagulation bath showed a post-spinning increase in the bath absorbance (Abs: 0.000 → 0.005), indicating a slight accumulation of the 4VBC-functionalised collagen product in the bath. A further negligible increase in solution absorbance (Abs: 0.000 → 0.007) was measured following the extrusion of the I2959-supplemented collagen dope (1 wt.% I2959) into the I2959-free coagulation bath. Taking into account the relatively low absorbance values (Abs < 0.1) and the potential for light interference [50], these UV-Vis readings suggest that a marginal fraction of functionalised collagen originally present in the dope is released to the coagulation bath, i.e. it does not contribute to fibre formation, alongside a small fraction of I2959 that is leached from the fibre-forming dope to the bath. The significant retention of the photoinitiator within the wet-spun fibres is likely attributable to the development of secondary interactions, such as hydrogen bonding or aromatic interactions, between 4VBC-functionalised collagen and I2959.

The minimal photoinitiator diffusion observed during the spinning process is also consistent with the improved handleability exhibited by the UV-cured collagen fibres (in comparison to their non-cured counterparts). This was mostly observed following coagulation in an ethanol bath supplemented with 0.5 wt.% I2959, whereby no increase in post-spinning bath absorbance was recorded (Table S3, Supp. Inf.), suggesting that the reduced photoinitiator concentration gradient between the dope and the bath contributed delayed soluble factor diffusion away of the dope. These wet spinning conditions were therefore adopted for the fabrication of wet-spun collagen fibres and subsequent fibre characterisation.

### 3.3 Fibre morphology

Following confirmation of the chemical composition, attention was directed towards investigating the fibre morphology in the wet-spun samples. Optical microscopy revealed the formation of homogeneous individual fibres in both the UV-cured and drop-cast states (Figure 2A-D), with fibre diameters of 173 ± 52 µm (n = 320) and 161 ± 56 µm (n = 352), respectively. A linear density of 5 ± 1 tex was measured for fibres in the natural relaxed state.

Fibre loading with BTB resulted in slightly smaller, though statistically insignificant, fibre diameters, likely due to hydrogen bonding effects following drop casting and solvent evaporation. In contrast, larger fibre diameters were observed following sample incubation in aqueous environments, consistent with the water-induced swelling behaviour of type I collagen.

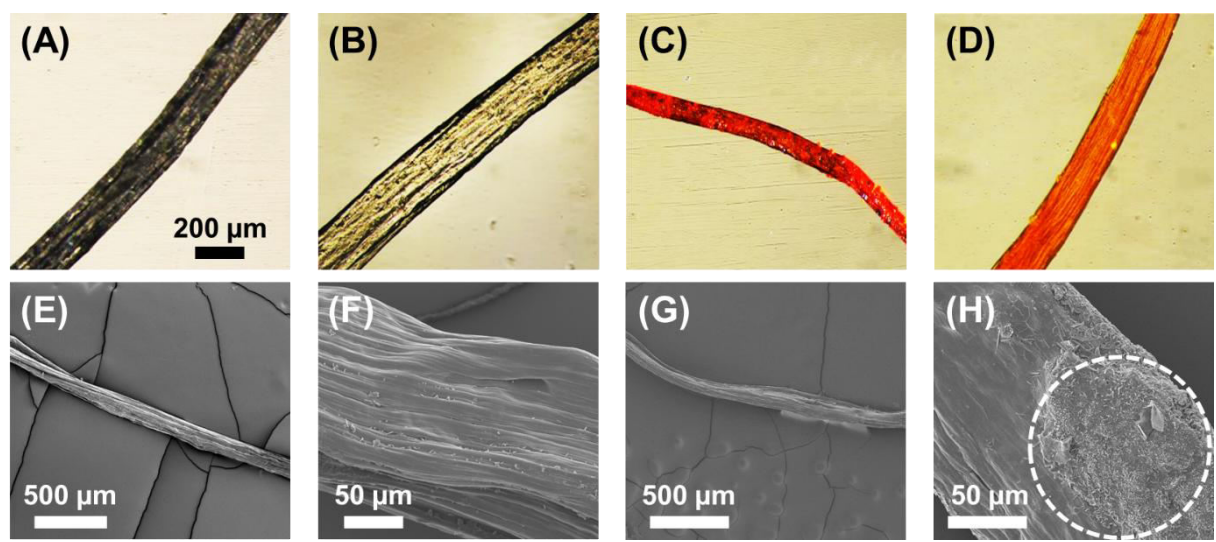

**Figure 2**. Morphology of UV-cured wet spun collagen fibres. (A-D): Optical images of BTB-free (A-B) and BTB-loaded (C-D) fibres in the dry (A, C) and hydrated (B, D) states. Scale bar applies to all images. (E-H): Electron microscopy images of BTB-free (E-F) and BTB-loaded (G-H) fibres. The dashed white circle highlights a localised BTB drop cast region of reduced fibre diameter, most likely arising from solvent evaporation effects following drop-casting.

Closer inspection of the fibre morphology by SEM revealed a longitudinally striated fibre structure (Figure 2E-H), attributed to a relatively high viscosity of the collagen wet-spinning solution [51], and the formation of physical interactions between collagen-grafted 4VBC residues [46], ultimately generating uneven shear forces during the wet-spinning process [52]. The dimensional variations observed in drop-cast samples by light microscopy were corroborated by SEM (Figure 2H), where local reductions in fibre diameter were detected, most likely arising from the effect of solvent evaporation following drop casting and molecular reorganisation of collagen molecules via hydrogen bonding.

Overall, these investigations demonstrate the manufacturability of 4VBC-functionalised collagen into textile fibres. Nonetheless, further optimisation of the wet-spinning parameters and process remains necessary to achieve improved control over fibre morphology, fineness, and length.

### 3.4 Thermal behaviour

Previous investigations of chemical composition and fibre morphology were subsequently complemented by thermal analysis, aimed at assessing the impact of BTB loading on the thermal properties of the wet-spun collagen fibres. TGA indicated increased thermal stability for the BTB-loaded (B-4VBC*), in comparison to BTB-free (F-4BC*), fibres (Figure 3A).

The BTB-loaded samples exhibited a remaining mass of 20 ± 2 wt.% at 800°C, compared to <1 wt.% in the BTB-free samples. Compared to F-4VBC*, the drop-cast fibres displayed more rapid mass loss below 230°C, while enhanced mass retention was observed at higher temperatures. Upon comparison of the TGA profiles beyond water evaporation (T > 100°C), the first linear decrease in mass was noted in the range of 250-350°C for F-4VBC* (mass loss ≈ 30 wt.%), comparable to other type I collagen samples [53]. In contrast, this transition occurred at a higher temperature and within a narrower temperature range in the drop-cast samples (T: 300-350°C), whereby approximately half of the mass loss (≈ 16 wt.%) was recorded relative to F-4VBC*. As the decomposition temperature of BTB has been reported to be approximately 230°C [54], these observations support the encapsulation of BTB within the fibres, in agreement with the preceding FTIR (Figure 1) and gravimetric loading efficiency results.

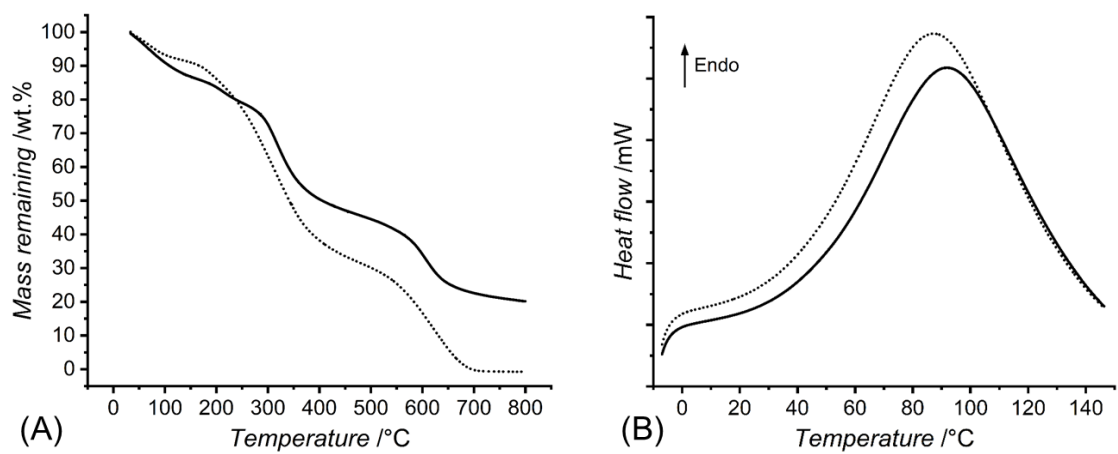

**Figure 3.** Representative TGA (A) and DSC (B) thermograms of BTB-loaded (solid line) and BTB-free (dotted line) UV-cured wet-spun collagen fibre.

In addition to TGA, DSC thermograms of dry BTB-free and BTB-loaded fibres revealed a single endothermic transition at 86 ± 5°C and 91 ± 6°C, respectively (Figure 3B and Table 1). This observation is consistent with previous reports on dry covalently crosslinked type I collagen [55] and is attributed to the denaturation of collagen triple helices into random coils [40, 46, 56]. This result further supports the presence of collagen triple helices within the wet-spun fibres, while the slight, though statistically insignificant, increase in denaturation temperature ($T_d$) in the dry samples of B-4VBC* samples compared to variants F-4VBC* correlates with the formation of secondary interactions between BTB and the collagen network at the molecular level.

**Table 1.** Thermal, swelling, and wet-state mechanical properties of BTB-free and BTB-loaded fibres determined by weight, DSC and tensile measurements, respectively. $T_d$: denaturation temperature; SR: equilibrium swelling ratio in distilled water; UTS: ultimate tensile strength; $\varepsilon_b$: elongation at break; E: tensile modulus; G: gel content. Three replicates were employed for each measurement except for tensile testing, whereby five replicates were measured for each sample. N/A: not applicable.

| Sample ID | $T_d$ (°C) | SR (wt.%) | UTS (MPa) | $\varepsilon_b$ (%) | E (MPa) | G (wt.%) |
|---|---|---|---|---|---|---|
| F-4VBC*  | 86 ± 5 | 492 ± 73 | 15 ± 7 | 14 ± 4 | 232 ± 32 | 81 ± 6 |
| F-4VBC*B | 91 ± 6 | 323 ± 79 | 12 ± 3 | 17 ± 5 | 265 ± 51 | N/A |

## 3.5 Characterisation of physical properties

Following the characterisation of the dry fibres, attention turned to sample testing in aqueous environments, with the aim of assessing macroscopic properties and infection responsivity in use-inspired conditions. The mean swelling ratio (SR) recorded at equilibrium after two hours of incubation in deionised water was 492 ± 73 wt.% and 323 ± 79 wt.% for BTB-free and BTB-loaded fibres, respectively (Table 1). The slightly lower, though statistically insignificant, SR observed following fibre encapsulation with the dye aligns with the formation of secondary interactions between the dye and the collagen network, as indicated by the slight increase in $T_d$ (Figure 3B) and the appearance of an additional shoulder peak in the FTIR spectrum (Figure 1) of the BTB-loaded samples compared to their BTB-free counterparts.

The aforementioned SR values exhibited by the fibres were significantly lower than those recorded when the same material was processed into cast films [31], which displayed an equilibrium swelling ratio of 1915 ± 311 wt.% under identical experimental conditions. To investigate the origin of this discrepancy, the gel content (G) of the fibres was quantified as an indirect measure of the crosslink density. G values were recorded in the range of 79 ± 6 and 92 ± 1 wt.% in samples wet spun into ethanol baths supplemented with varying concentrations of I2959 (Figure 4).

The gel content data revealed no statistical significance among samples, indicating that the supplementation of the coagulation bath with I2959 had a negligible effect on the crosslink density of the resulting UV-cured wet spun fibres.

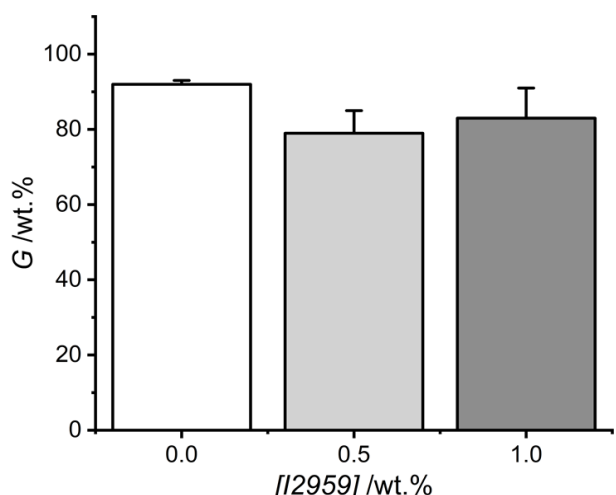

**Figure 4.** Gel content (G) exhibited by collagen fibres (n=3) wet spun from an I2959-supplemented collagen dope (1 wt.% I2959) in an ethanol bath supplemented with increasing concentration of I2959 (0—1 wt.%) and UV-cured. No significant differences were measured between the three sample groups.

This finding indirectly supports earlier UV-Vis absorbance data, which suggested photoinitiator retention within the wet spun fibre following dope extrusion. These UV-Vis data also suggest that the fraction of I2959 released to the coagulation bath did not seem to be inversely correlated with the observed gel content variation of the resulting UV-cured fibres This observation possibly hints at the fact that decreased concentration of I2959 in the wet spinning dope compared to the one used in this study (0.5 wt.% I2959) may still be sufficient to accomplish comparable crosslink density.

Although no statistically significant differences were observed, the relatively high G values align with earlier reports for comparable cast films [31], suggesting similar crosslink densities and swelling behaviours between the two material formats. However, the significantly reduced SR observed in the fibres compared to their thin film counterparts' points to additional structural factors influencing water uptake. The most logical explanation for the markedly lower SR observed in the fibres, as compared to the films, lies in the wet spinning process itself. During extrusion into the coagulation bath, collagen molecules experience shear and elongation flow, promoting longitudinal alignment and tight molecular packing of the collagen molecules along the fibre axis [46, 51, 57]. This structural organisation limits the formation of a porous microstructure, thus reducing the free volume and the capacity of the fibres to absorb water. In contrast, cast films allow molecules to adopt a more relaxed and isotropic structural arrangement prior to crosslinking, preserving greater porosity and water-accessible swelling domains. These observations highlight the role of processing-induced microstructure in dictating the functional properties of collagen-based biomaterials.

Despite their significantly reduced swelling compared to thin films, the SR values recorded for the fibres remain comparable to those exhibited by commercial polyurethane-based wound dressings designed to maintain a moist wound environment [43, 58]. In this context, the fabrication of individual collagen-based fibres is appealing aiming to meet current industrial practice for the manufacture of fibrous wound dressings, whereby fibre bundles are twisted into yarns and subsequently woven into a three-dimensional porous textile construct. This structural configuration can be beneficial aiming at enhanced absorption and retention of biological fluids relative to individual fibres. On the other hand, the relatively low SR of the individual fibres may prove advantageous in applications such as resorbable sutures [59].

In addition to the swelling tests, the enzymatic degradability of the wet-spun fibres was assessed, given the rapid cleavage of collagen-based materials *in vivo*. To explore this challenge, UV-cured wet spun fibres were prepared from clinically approved bovine type I collagen and incubated in a collagenase-rich medium (2CDU·mL$^{-1}$) for up to 72 hours, followed by ethanol series dehydration [42, 43]. Individual fibres retrieved following 24-hour incubation indicated around one third of relative mass ($\mu_{rel}$= 34±14 wt.%), while further mass decrease was observed at 48 hours ($\mu_{rel}$= 11±7 wt.%) and 72 hours ($\mu_{rel}$= 11±4 wt.%) (Table S4, Supp. Inf.). These trends were confirmed by the rat tail collagen-based variants of wet spun fibre (Figure 5), while complete degradation of native rat tail collagen was observed within 24 hours, in line with the absence of a UV-cured covalent network at the molecular scale. Interestingly, similar values of $\mu_{rel}$ were also measured with rat tail collagen thin films, indicating comparable degradation rate of the UV-cured collagen network independently of its processing into fibre or film and respective material size (Figure 5).

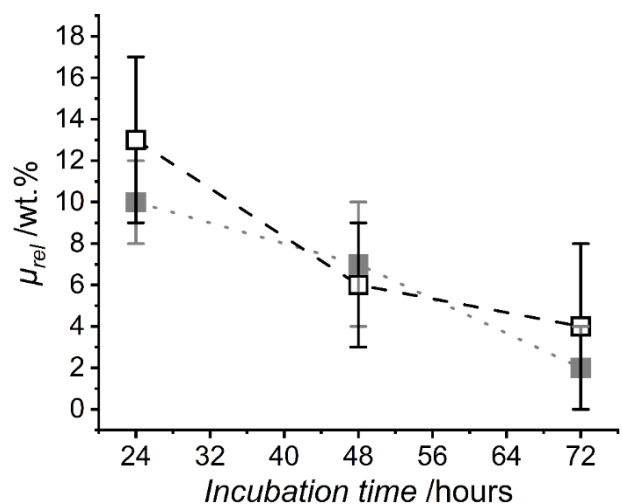

**Figure 5.** Temporal profiles of relative mass ($\mu_{rel}$) exhibited by samples (*n=3*) of UV-cured wet spun fibre (—■—) and thin film (···□···) following 72-hour incubation in a collagenase-rich medium (2 CDU; pH 7.4; 37°C). Data are presented as mean ± standard deviation; no significant differences were measured between the two sample groups at each time point.

The aforementioned relative mass values revealed by the UV-cured fibres indicate enhanced enzymatic stability compared to commercially available collagen-based membranes and previously reported photocured collagen prototypes. Vallecillo et al. found that Fibro-Gide, Mucograft, and Mucoderm were completely degraded within 48 hours under experimental conditions comparable to the ones reported in this study [60]. Similarly, Ali et al. found that UV-cured methacrylated collagen hydrogels lost over 50% of their mass after only 4 hours in a collagenase-supplemented solution (2.5 CDU) [61]. Although the degradation times observed in the present study appear relatively short, it is well established that degradation *in vivo* typically proceeds more slowly than *in vitro*, as physiological conditions are generally less aggressive than the idealised enzymatic environments used in laboratory assays [62].

While the materials tested in this degradation study were completely submerged in the enzymatic medium, further research is warranted to develop clinically relevant degradation models to improve *in vitro* simulations for the *in vivo* applications of the collagen fibres. Degradation studies in these use-inspired environments should include tensile testing at various degradation time points to assess the effect of degradation on mechanical performance and structural integrity [63]. In this context, ensuring material integrity stability over time is essential to reduce risks of wound exposure to exogenous bacteria. However, achieving controlled degradation of collagen is a desirable property to foster wound healing in hard-to-heal wounds, whilst also removing the need for secondary surgery in suture applications [32, 64].

**3.6 Dye Release and colour-change capability**

Following confirmation of dye loading efficiency and swelling properties, the colour-change behaviour of the wet-spun collagen fibres was assessed in simulated wound fluids. The fibres were incubated in acidic (pH 5) and alkaline (pH 8) McIlvane buffer solutions for 96 hours at room temperature to determine whether dye leaching would occur during a clinically relevant wound dressing application period, as this could irreversibly alter the wound environment and compromise dressing longevity and performance. Testing *in vitro* was conducted at room temperature to reflect the conditions experienced at the wound–dressing interface, which is typically cooler than core body temperature [65]. The temperature of the chronic wound bed has been reported to be around 33°C, i.e. lower than 37 °C, and to be as low as 25.3°C following dressing change. Buffer solutions were adjusted at pH 5 and pH 8 aiming to investigate the pH responsivity of the collagen samples in the most extreme healing and chronic wound environments, respectively, and demonstrate a clear, visually detectable pH-induced colour change. Incubation studies were therefore conducted in the above conditions aiming to investigate the extent of dye retention within the fibres, as a crucial aspect to ensure durable infection responsiveness and to minimise the risk of dressing-induced toxicity.

Following complete submersion of the fibres in acidic McIlvaine buffer (pH 5), negligible dye release was detected in the supernatant, corresponding to a mean value of 12 wt.% after 96 hours (Figure 6A and C). The significant retention of the dye in this aqueous environment is most likely due to the development of electrostatic interactions between BTB and the collagen-based fibres (Scheme 1). At pH 5, the negatively charged sulfonate groups of BTB can interact with the protonated amines of the collagen network (p$K_a$ ≈ 9), consistent with the minimal release previously observed under these conditions [38]. These electrostatic interactions likely complement any $\pi-\pi$ aromatic stacking interactions between BTB and 4VBC-crosslinked collagen molecules, as previously supported by FTIR data (Figure 1).

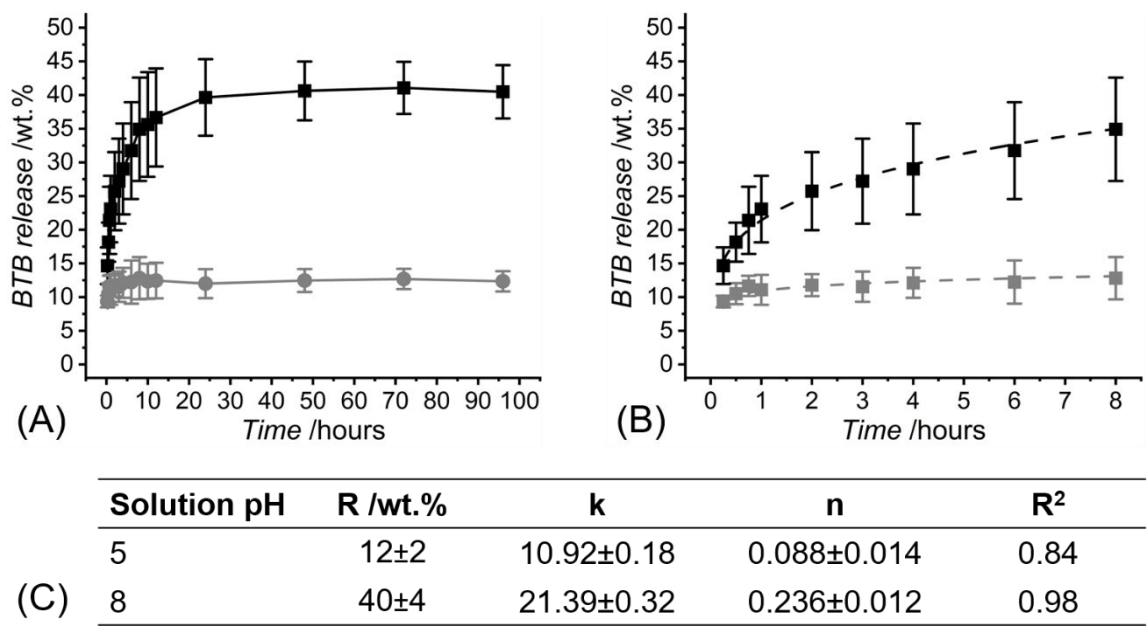

**Figure 6.** BTB release measurements in acidic (pH 5, grey) and alkaline (pH 8, black) McIlvaine solutions. (A): Release profiles revealed by the BTB-loaded fibres following 96-hour incubation. Lines are guidelines to the eyes. (B): Fitting of the first 60% release data according to Korsmeyer-Peppas model (y= k · x$^n$). (C): values of the 96-hour release (R), Korsmeyer-Peppas model coefficients (k, n) and coefficient of determination ($R^2$) determined following incubation of drop cast fibres in either acidic (pH 5) or alkaline (pH 8) McIlvaine solutions. Data are presented as mean ± standard deviation (*n=3*).

| Solution pH | R /wt.% | k | n | $R^2$ |
|---|---|---|---|---|
| 5 | 12±2 | 10.92±0.18 | 0.088±0.014 | 0.84 |
| 8 | 40±4 | 21.39±0.32 | 0.236±0.012 | 0.98 |

In contrast to the acidic environment, a significant increase in dye release was observed following 96-hour incubation in alkaline McIlvaine solution (pH 8), rising from 12 ± 2 wt.% to 40 ± 4 wt.% (Figure 6B and C). This increase is likely due to the deprotonation of collagen amine groups under alkaline conditions and the resulting electrostatic repulsion of the negatively charged sulfonate groups of BTB and the now neutral or negatively charged collagen network (Scheme 1).

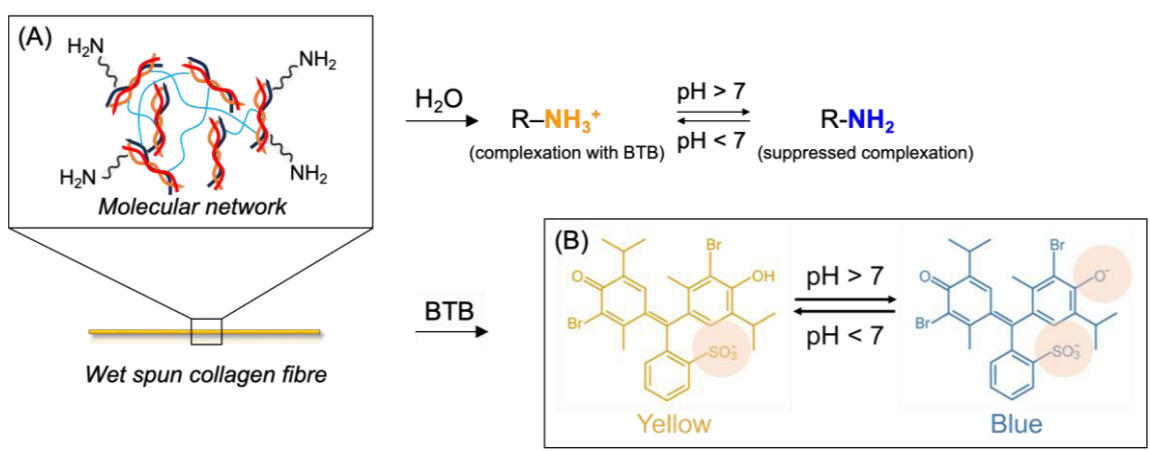

**Scheme 1.** Proposed complexation mechanism underpinning the retention of BTB in the wet spun collagen fibre in aqueous environments. An increase in solution pH generates reversible protonation of the free amino groups of the UV-cured collagen network (A) and molecular reconfiguration of BTB towards an increased negative charge, as highlighted by the corresponding sulfonate and phenolate groups (B). Consequently, varying degree of electrostatic complexation between the dye and the collagen network can be expected.

To gain further insight into the release mechanism and test the aforementioned hypothesis, the release data were fitted using the Korsmeyer-Peppas model (Figure 6B). The model showed a strong correlation with the release profile observed under alkaline conditions (pH 8), as supported by an $R^2$ value of 0.98. The release exponent, *n*, was measured to be below 1, with a value of 0.236, consistent with a diffusion-controlled release mechanism. In contrast, poorer model fit was observed when applying Korsmeyer-Peppas model to the release data recorded under acidic conditions (pH 5), yielding $R^2$ and *n* values of 0.84 and 0.088, respectively. While the low *n* value still indicates a diffusion-driven mechanism, the reduced value of $R^2$ is likely attributable to the relatively low dye release ([BTB] = 12 ± 2 wt.%) after 96 hours in the acidic medium.

While the aforementioned testing was carried at room temperature, rather than wound bed temperature (T≈ 33 °C), it is unlikely that a temperature increase in this range would significantly affect the dye release capability of the presented collagen materials, given the presence of a crosslinked network [66] and the lack of thermosensitive segments [67] at the molecular scale.

In our previous work, a two-layer mesh composed of electrospun BTB-encapsulated poly(methyl methacrylate-co-methacrylic acid) (PMMA-co-MAA) fibres displayed BTB release levels of up to 99 wt.% after two hours of incubation in alkaline conditions, while a markedly lower release of 26 wt.% was recorded at pH 5 after 96 hours [68]. This high release correlates with the absence of ionisable primary amine groups in the PMMA-co-MAA backbone, contrasting with the significantly reduced release observed from the drop-cast UV-cured wet-spun collagen fibres developed in the present study. On the other hand, curcumin-loaded chitosan films were found to release up to ca. 65 wt.% of curcumin following 72 hours of incubation in aqueous environments. This fraction was reduced to around 40 wt.% when the chitosan films were encapsulated with curcumin-loaded silica nanoparticles [69]. Here, lower curcumin release (≈ 30 wt.%) was measured in highly acidic solutions (pH 2) compared to the case of mildly acidic (pH 6) and slightly alkaline (pH 7.4) conditions, an effect attributed to protonation of the primary amino groups of chitosan in acidic media. Similar levels of anthocyanin release (≤ 45 wt.%) were also observed when a two-layer composite of agar and carrageenan encapsulated with anthocyanin-loaded liposomes was incubated in aqueous solutions for 70 minutes, while complete release was recorded in the absence of liposomes [70]. In contrast to these reports, the drop casting method employed in this study offers a compelling, low-cost and effective strategy for integrating wet-spun fibres with BTB retention capability in both acidic and alkaline environments, aiming to accomplish long-lasting halochromic capability for wound monitoring applications, without the need for particles or complex microstructures. The resulting high loading efficiency and supporting evidence of BTB-collagen electrostatic interactions suggest that this drop casting protocol also allows for the loading of tailored BTB dosages with minimal risks of abrupt, unwanted dye release and no cytotoxic effects from the UV-cured collagen materials [38, 43, 46, 71].

While the encapsulation of pH-sensitive dyes prior to fibre production can alter polymer processability and lead to variations in fibre morphology and microstructure, the drop-casting method employed in this study enabled consistent loading of BTB onto the UV-cured hydrogel networks, whether in the form of wet-spun fibres or cast films [31]. The release profiles recorded with the drop cast fibres display no significant difference in alkaline conditions (pH 8; 12 h) in comparison to the thin films made of the same material, unlike the case when the same samples were incubated in acidic conditions (pH 5; 12 h). We hypothesise that the higher release observed from the fibres, as compared to the films, is due to the tightly packed and aligned molecular architecture of the wet-spun fibres, which may hinder dye absorption into the fibre core. The difference in material geometry may also contribute to this effect, given the microscale diameter of the wet-spun fibres relative to the millimetre-scale diameter of the collagen thin films.

Having confirmed dye release profiles in wound-relevant aqueous environments, attention was focused on assessing the colour-change functionality of the resulting BTB-loaded wet-spun collagen fibres (Table 2). The native fibres prior to drop casting exhibited a slight yellow hue (L = 68, C = 34, h = 87). Following dyeing with an aqueous BTB solution, a notable colour shift was recorded (ΔE ≈ *27*), the colour became richer (C = 52) and slightly darker (L = 57), indicating substantial dye uptake and strong visible colouration, consistent with previously reported loading efficiency. After 96 hours of incubation in acidic McIlvaine buffer (pH 5), only a moderate shift in the LCh colour coordinates was observed (ΔE ≈ 17), with hue and chroma largely preserved. This limited ΔE value suggests that the BTB remained mostly entrapped within the fibre matrix and retained its halochromic activity, indicating good dye robustness under prolonged aqueous conditions.

**Table 2.** Lightness (L), chroma (C), and hue (h) values of dry samples of BTB-free fibres (F-4VBC*) and BTB-loaded fibres (B-4VBC*), as well as dry samples of BTB-loaded fibres following 96-hour incubation in a McIlvaine solution at pH 5 and pH 8. The total colour difference (ΔE) is shown relative to the BTB-loaded fibre. N/A: not applicable. Data are presented as mean ± standard deviation (n=3).

| Sample ID | L | C | h | ΔE |
| --- | --- | --- | --- | --- |
| F-4VBC* | 68 | 34 | 87 | 27 |
| B-4VBC* | 57 | 52 | 70 | N/A |
| B-4VBC* (96 h, pH 5) | 43 | 52 | 61 | 17 |
| B-4VBC* (96 h, pH 8) | 41 | 20 | 131 | 70 |

In contrast, immersion in alkaline McIlvaine solution (pH 8) resulted in a pronounced colour change (ΔE ≈ 70), shifting from yellow-orange to blue-green. This *ΔE* value approaches the theoretical maximum perceptual difference (ΔE ≈ 100) in the LCh colour space. This signifies a profound and unambiguous colour transition, far more than would occur from incidental fading, therefore making it highly detectable, even by non-experts. This substantial shift is consistent with the halochromic response of the embedded BTB and demonstrates clear pH-responsiveness of the fibre construct under alkaline conditions.

This colour change was demonstrated to be reversible and robust over clinically relevant time periods (Figure 7). The fibres appeared blue-green under alkaline conditions and consistently reverted to yellow-orange after multiple transfers into the acidic solution. Although the colour change capability of the fibres was demonstrated at room temperature only, it is likely that comparable behaviour would be observed at wound bed temperature, given the excellent thermal stability and consistent colour change capability of BTB at a temperature range of 25–90 °C [72].

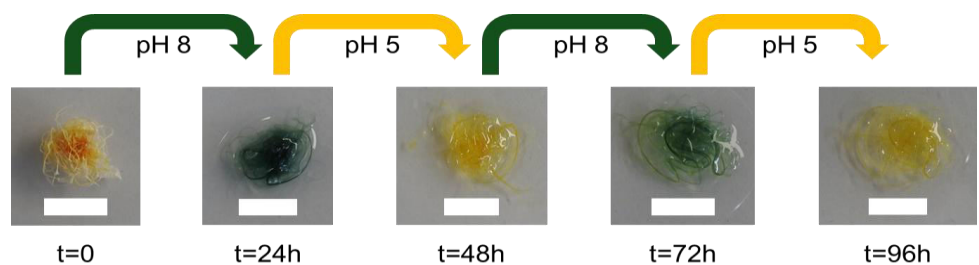

**Figure 7.** Photographs of BTB drop-cast collagen fibres displaying reversible colour change and durable infection-responsive capability over 96 hours following alternating incubation steps in acidic (pH 5) and alkaline (pH 8) McIlvaine solutions. *Scale bars 10 mm*.

This pronounced and repeatable colour change under alkaline conditions highlights the suitability of the material for infection detection. This would allow clinicians and patients alike could visually monitor wound alkalinisation, a known early infection indicator [73] without the need for specialist instruments. The strong ΔE ensures robustness against changes in ambient lighting or moisture variability and supports potential smartphone-based or naked-eye monitoring of local pH.

Future work on colour measurements could encompass a broader range of pH values to yield a comprehensive spectral profile. The integration of these calibration datasets with smartphone-based colour analysis or dedicated mobile applications offers the potential for rapid, continuous and objective pH readouts at either the point of care or remotely. This methodology serves to complement the naked-eye monitoring demonstrated in the current feasibility study, aiming towards enhanced precision, reduced user subjectivity, and improved support for remote or self-managed wound monitoring.

**3.7 Wet-state tensile properties of UV-cured wet spun fibres**

To evaluate whether the material retained mechanical competence in physiological environments, tensile testing was performed on both BTB-free and BTB-loaded collagen fibres after they reached equilibrium in PBS (Figure 8). No statistically significant differences were observed in ultimate tensile strength (UTS), elongation at break ($\varepsilon_b$), or tensile modulus (E) between the two groups (Table 1), whose average values were measured as 12-15 MPa, 14-17 %, and 232-265 MPa, respectively. These findings suggest that any secondary molecular interactions between BTB and the collagen matrix did not translate into a measurable impact on bulk mechanical properties.

As shown in Figure 8, the stress-strain curves of both BTB-free and BTB-loaded fibres displayed a similar profile, which was characterised by an initial, higher-stiffness, linear elastic region, followed by a transition to a lower-stiffness region. This yielding behaviour is likely associated with structural rearrangements, such as alignment of the collagen molecules in the direction of applied load or localised yielding of discrete domains [74, 75]. The extent of yielding remains relatively limited, as evidenced by the modest deviation from linearity.

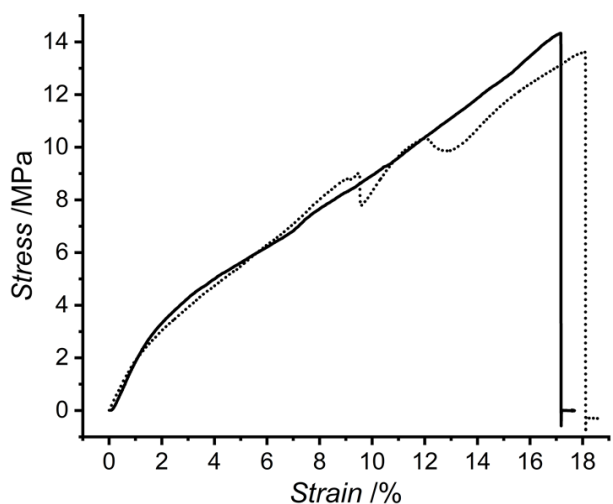

**Figure 8.** Representative stress–strain curves (n=5) from uniaxial tensile tests performed on individual, UV-cured collagen fibres, either BTB-free (dotted line) or BTB-loaded (solid line), wet spun in a photoactive ethanol bath containing 0.5 wt.% I2959. All fibres were incubated for 2 hours in PBS (10 mM, pH 7.4) prior to testing.

The subsequent lower-stiffness linear region may reflect the response of more compliant, hydrated zones within the covalently crosslinked collagen network.

Comparable manufacturing strategies have been reported in the literature to produce wet-spun collagen fibres with similar tensile strength and moduli under hydrated conditions. Notably, Pedro Kato et al. reported that fibres crosslinked via glutaraldehyde or through dehydration and cyanamide treatment exhibited enhanced tensile performance [76]. Glutaraldehyde-crosslinked fibres showed slightly higher tensile strength and modulus compared to those produced in this study, whereas comparable properties were measured with cyanamide-treated fibres (UTS: 24-31 MPa; E: 170-200 MPa). In another study, Tonndorf et al. developed wet-spun collagen yarns using a 6-hole spinneret followed by glutaraldehyde crosslinking, reporting a linear density of ca. 33 tex, a UTS of 40 ± 4 MPa, and a Young's modulus of 281 ± 15 [77] In contrast, the UV-cured wet-spun developed in this study exhibited a substantially lower linear density ($\lambda \approx 5$ tex), indicating superior mechanical integrity per unit mass.

Collagen extruded fibres have also been crosslinked via thermal dehydration at different temperatures (60, 80, 110, and 140 °C) and time intervals (1, 3, and 5 days). The crosslinked collagen fibres showed noticeably high UTS and elastic moduli in the dry state, exceeding 600 and 8000 MPa, respectively [78]. The magnitude of mechanical strength did not translate when fibres were hydrated, whereby a maximum value of UTS and moduli of ca. 92 and 90 MPa was observed, respectively. This value of UTS proved to be higher in comparison to the one measured in this study, an observation that is attributed to the significantly decreased fibre diameter than reported in this study. Significantly higher UTS was also recorded in collagen fibres crosslinked via either carbodiimide chemistry (UTS≈ 40 MPa) or glutaraldehyde (UTS≈ 138 MPa) [79], likely due to the application of fibre-drawing post-spinning.

Other than wet spinning, multifilament fibres were fabricated from blends of collagen and poly(ethylene oxide) (PEO) via contact drawing [80]. The reported mechanical properties (UTS= 24.79±4.07 MPa; E= 344.3±0.0 MPa) were comparable to those measured for the individual UV-cured collagen fibres developed in this study, despite the relatively modest swelling properties exhibited by the former samples.

The combination of remarkable wet-state tensile strength and reduced linear density supports the suitability of the UV-cured wet spun fibres as building blocks of textile-based wound dressings and highlights their suitability for use as lightweight, resorbable sutures. For instance, polypropylene 6-0 (*PP*) sutures have been reported to display a linear density comparable to that of the fibres produced in this study, but with significantly higher tensile modulus (E $\approx$ 20 GPa) [81]. However, *PP* fibres lack biodegradability in physiological environments, necessitating the need for surgical removal following wound closure [81, 82]. In contrast, the collagen fibres investigated here offer enzymatic degradability, presenting an appealing, patient-friendly alternative, while their tensile strength equips them with knot-forming and knot-tying abilities, further supporting their potential for suture applications (Section 3.8).

### 3.8 Fibre manufacturability onto woven dressings and resorbable sutures

Wet spinning of individual fibres is a well-established manufacturing process routinely employed by the textile industry for producing fibre-based healthcare materials such as wound dressings and implantable sutures. It was therefore of interest to investigate the scalability of the manufacturing route developed here relative to industrial requirements.

Optimisation of our wet spinning apparatus led to a semi-automated extrusion setup, whereby the use of a 0.75 m coagulation bath allowed for the fabrication of fibres of up to 2.4 m in length (Figure S3, Supp. Inf.). Due to the reduced mechanical integrity of freshly spun fibres, the extruded fibres were left in the coagulation bath for up to 10 minutes to enable complete collagen coagulation and minimise risks of fibre damage, prior to fibre collection onto the motorised spindle and UV curing (Scheme 2).

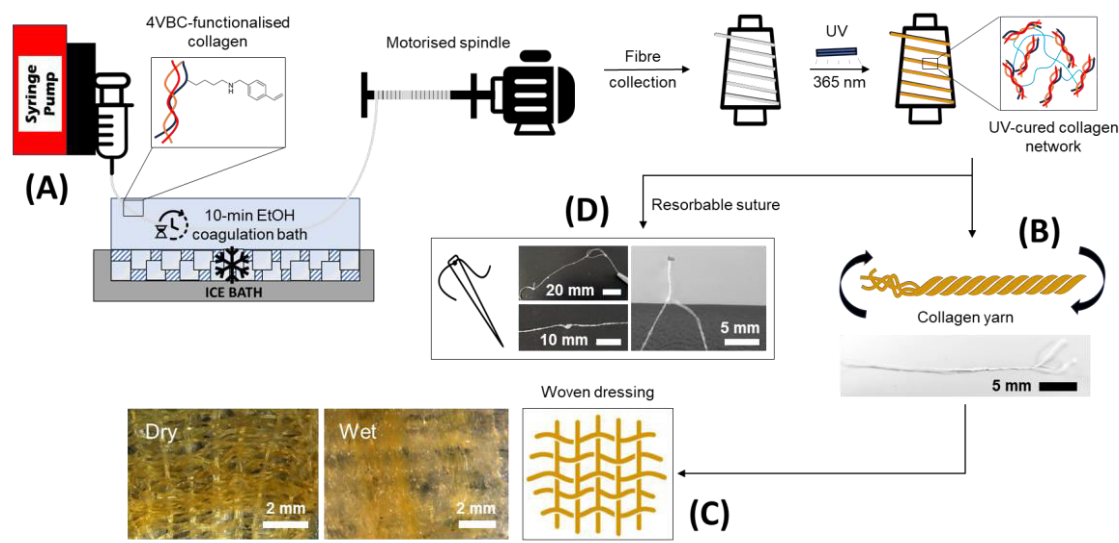

**Scheme 2.** Scalable wet spinning of individual collagen fibres with enhanced mechanical competence. Fibres are extruded in an ethanol (EtOH) bath in contact with ice (A), and collected onto a motorised spindle, prior to UV curing and ethanol series dehydration. Resulting fibres can be twisted into yarns (B) to generate woven dressings (C) or used as resorbable suture with knot-forming and knot-tying abilities (D). The infection-responsive colour-change capability of the resulting products enables prompt infection diagnosis and tailored standards of care

The application of low temperature (T= -20 °C) during fibre extrusion proved key to generate uniform fibres and minimise risks of collagen denaturation, as evidenced by FTIR analysis (Figure 1). At the same time, the elongated shape of the coagulation bath enabled the reproducible manufacture of fibres with scalability to industrial manufacturing. The employment of a wet spinning dope with increased collagen concentration, together with an UV light system with improved spectral alignment, could provide an additional experimental space to further process automation and accomplish rapid in situ curing, damage-free fibre drawing and spooling. Nevertheless, the resulting UV-cured wet-spun fibres displayed sufficient mechanical competence to be successfully twisted into yarns and subsequently woven into a dressing prototype, in line with current industrial manufacturing practices [83, 84]. The synthesis of a UV-cured molecular network proved therefore crucial for improving yarn compliance and minimising damage during handling, ultimately yielding a collagen-based textile dressing with integrated infection-responsiveness via colour-changing capabilities. Given the demonstrated fibre manufacturability, future work would be needed to investigate the BTB release profiles and pH-induced colour change capability described by the woven dressing prototype in use-inspired environments. Testing of the woven fabric will also be relevant to determine whether the larger volume of the corresponding collagen yarns (with respect to the wet spun fibres) leads to an increased retention of BTB.

Beyond their application in wound dressings, the elasticity and fineness of the UV-cured fibres supported their suitability as resorbable sutures, as evidenced by their ability to form and tie knots comparably to the case of commercial silk-braided sutures (Figure S4, Supp. Inf.). For this application, however, further research is needed to assess the biodegradability of the fibres in wound-simulated environments, whereby the material is applied on top of a simulated wound fluid layer rather than submerged in the solution, as carried out in this study. Ensuring delayed changes in mechanical properties during degradation will be essential to support the wound during the healing process and to minimise the risk of wound exposure to the external environment by degrading at the same rate as the wound healing process [63].

## Conclusions

This study presents the design of individual pH-responsive textile fibres as building blocks of smart wound monitoring devices capable of reversible colour change in response to infection-associated alkaline shifts (pH 5 → 7). Scalable wet spinning of chemically functionalised type I collagen molecules was successfully integrated with UV-induced network formation and drop casting of pH-responsive bromothymol blue. The resulting drop-cast collagen fibres displayed an average thickness of 173 μm and an average dye loading efficiency of 99 wt.%, enabling visual colour shifts when exposed to acidic (pH 5, ΔE= 17) or alkaline (pH 8, ΔE= 70) solutions. Wet spinning of the photoinitiator-loaded collagen dope in a 0.75 m coagulation bath yielded photoactive fibres of up to 2.4 m in length. Marginal increase in the coagulation bath absorbance following wet spinning (0.000 → 0.0007) indicated retention of the photoinitiator in the fibre-forming collagen jet, so that direct UV curing generated insignificant variations in gel content and mechanically competent fibres in the hydrated state. An average elongation at break, ultimate tensile strength, and tensile modulus of up to 17%, 15 MPa, and 265 MPa were recorded, respectively. Knots could be tied in individual fibres, fibres twisted into yarns, and yarns woven into textile dressing prototypes which have been successfully demonstrated *in vitro*. More than a threefold increase in mass was observed following incubation and water-induced swelling of BTB-loaded fibres *in vitro.* BTB was largely retained after 96 hours submersion in either acidic (pH 5) or alkaline (pH 8) solutions, corresponding to an average release of 12 wt.% and 40 wt.%, respectively. This retention is attributed to the development of electrostatic interactions between the dispersed negatively charged BTB molecules and the primary amino groups present in the UV-cured collagen network, in line with the increased retention capability of the fibres in acidic environments. Following 24 hours in a collagenase-rich medium, fibres retained up to ≈30 wt.% of their initial mass. This finding supports the potential use of these UV-cured fibres as resorbable sutures. In light of their mechanical performance, infection responsivity, and scalable manufacturability, these multifunctional fibres offer broad applicability as wound monitoring devices, enabling prompt infection diagnosis and tailored standards of care, while maintaining compatibility and durability in the wound environment.

## Acknowledgements

The authors gratefully acknowledge the financial support from Leeds Institute of Textile and Colour (LITAC) and the Clothworkers' Centre for Textile Materials Innovation for Healthcare (CCTMIH). The authors gratefully acknowledge LEMAS for their support and assistance conducting the SEM work. J.G. gratefully acknowledges the Whitworth Society for the financial support of his PhD program via the Senior Scholar Award. The authors would like to thank Mohammed Asaf (University of Leeds, School of Design) for technical assistance with TGA and DSC measurements.

## Conflict of interests

G.T. is named inventor on a patent related to the fabrication of collagen-based materials. He has equity in and serves on the board of directors of HYFACOL Limited.

Supporting Information

**An infection-responsive collagen-based wet-spun textile fibre for wound monitoring**

Jonathon Gorman,[1,2] Charles Brooker,[1,2] Xinyu Li,[1] Giuseppe Tronci[1,2]

[1] Clothworkers Centre for Textile Materials Innovation for Healthcare (CCTMIH), Leeds Institute of Textiles and Colour (LITAC), University of Leeds, United Kingdom

[2] Division of Oral Biology, St. James's University Hospital, School of Dentistry, University of Leeds, United Kingdom

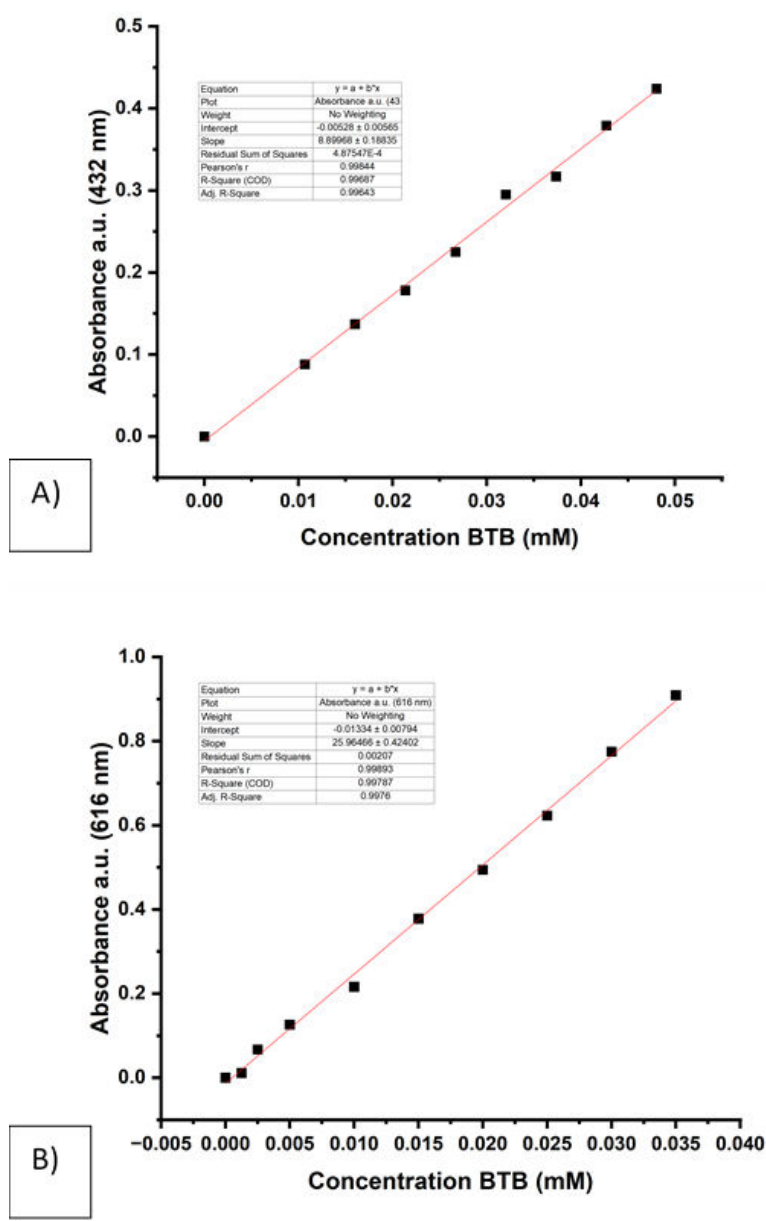

**Figure S1**. Calibration curves of BTB in McIlvane solutions adjusted to pH 5 (A) and pH 8 (B). Absorbance was measured at either 432nm (A) or 616nm (B). These calibration curves were used to quantify the release of BTB from the fibres in either acidic or alkaline environments.

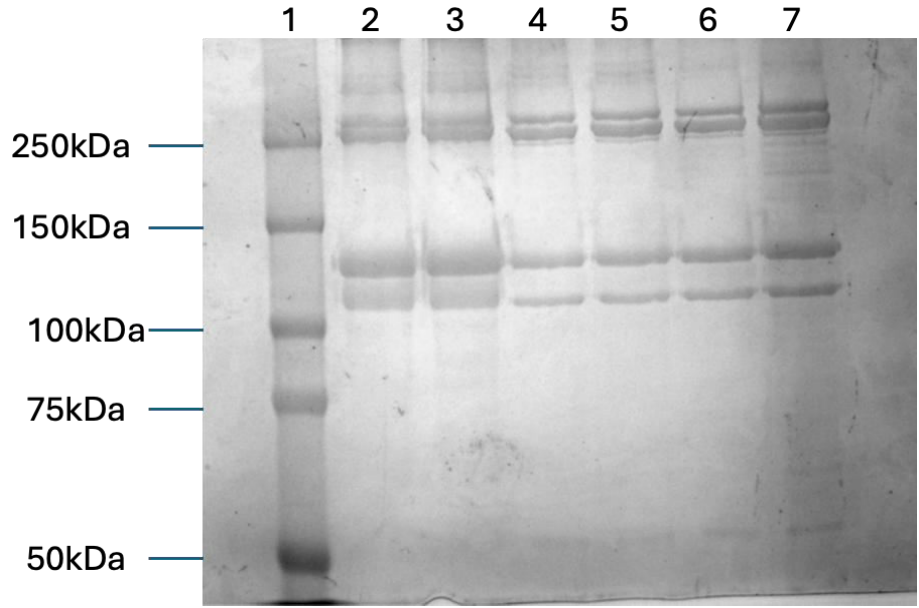

**Figure S2**. SDS-PAGE analysis of control (1), commercial bovine collagen (2-3, Collagen solutions™) in-house extracted rat tail collagen (4-7). All samples display the double alpha band (≈100 kDa) and beta band (≈250 kDa) characteristic of native type I collagen.

**Table S1.** Degree of functionalisation (F) measured via TNBS assay of three batches of 4VBC-functionalised rat tail collagen. The data are presented as average ± standard deviation (n=3).

| Batch | $F$ /mol.% |
|---|---|
| #1 | 35 ± 6 |
| #2 | 30 ± 7 |
| #3 | 34 ± 1 |

**Table S2.** Loading efficiency (LE) measured in the UV-cured wet spun collagen fibres by recording the dry mass of the samples prior to ($m_d$) and following ($m_f$) BTB drop casting. Average and standard deviation (S.D.) of LE are provided at the bottom of the table.

| Sample | $m_d$ /g | $m_f$ /g | BTB loaded /g | LE /wt.% |
|---|---|---|---|---|
| 1 | 0.01257 | 0.01266 | 0.0001 | 90 |
| 2 | 0.01103 | 0.01113 | 0.0001 | 100 |
| 3 | 0.01239 | 0.01249 | 0.0001 | 100 |
| 4 | 0.0076 | 0.0077 | 0.0001 | 100 |
| 5 | 0.00805 | 0.00825 | 0.0002 | 100 |
| 6 | 0.00923 | 0.00943 | 0.0002 | 100 |
| 7 | 0.00881 | 0.009 | 0.0002 | 95 |
| 8 | 0.00672 | 0.00678 | 0.00006 | 100 |
| 9 | 0.01344 | 0.0135 | 0.00006 | 100 |
| 10 | 0.01121 | 0.01127 | 0.00006 | 100 |
| 11 | 0.00456 | 0.00466 | 0.0001 | 100 |
| 12 | 0.00902 | 0.00922 | 0.0002 | 100 |
| 13 | 0.00947 | 0.00968 | 0.0002 | 105 |
| 14 | 0.01006 | 0.01025 | 0.0002 | 95 |
| | | | **Average ± S.D.** | 99±3 |

**Table S3**. Absorbance values of the coagulation bath recorded prior to (Abs$_{PRE}$) and post (Abs$_{POST}$) wet spinning. Data are presented as mean ± standard deviation (n=3). The wet spinning dope was prepared via dissolution of 4VBC (0.8 wt.%) and I2959 (0, 1 wt.%) in 17.4 mM acetic acid. The coagulation bath consisted of ethanol supplemented with varied concentration of I2959 (0–1 wt.%).

| [I2959] /wt.% | | Abs$_{PRE}$ /a.u | Abs$_{POST}$ /a.u | Mean difference |
| --- | --- | --- | --- | --- |
| dope | bath | | | |
| 0 | 0 | 0 | 0.0053 ± 0.0032 | -0.005* |
| 1 | 0 | 0 | 0.007 ± 0.0046 | -0.007* |
| 1 | 0.5 | 0.303 ± 0.0015 | 0.303 ± 0.0006 | 0 |
| 1 | 1 | 0.493 ± 0.002 | 0.483 ± 0.003 | 0.01** |

* Negative values of mean absorbance indicate diffusion of soluble factors from the fibre-forming wet spinning jet to the coagulation bath.

** Positive values of mean absorbance indicate diffusion of soluble factors from the coagulation bath towards the fibre-forming wet spinning jet, so that no release can be quantified.

$$50\ UI/ml = 400 \mu g/mL \therefore 2UI/mL = 16 \mu g/mL \quad \textbf{Equation S1}$$

$$2\ CDU = \frac{5mg}{312.5mL} = 16 \mu g/mL \quad \textbf{Equation S2}$$

Equations S1 and S2 were used in the degradation study *in vitro* to calculate and compare clinically relevant collagenase concentrations [1, 2] according to a collagenase activity of 125 CDU·mg$^{-1}$. CDU and UI (or U) are the same unit in this context as both undertake the same measurement for enzymatic activity. 1 CDU/UI is defined as the amount of enzyme that liberates 1 µmol of L-leucine following 5–hour incubation of collagen *in vitro* (37°C, pH 7.5) [3, 4].

**Table S4.** Relative mass ($\mu_{rel}$) measured during 72-hour incubation of drop cast bovine and rat tail collagen fibres and rat tail films (n=3) in a collagenase-supplemented TES buffer (pH 7.4, 37°C; 2 CDU). Average and standard deviation (S.D.) of $\mu_{rel}$ is reported following 24, 48 and 72 hours of sample incubation.

**Bovine collagen-based fibre degradation**

| 24 h | $m_d$ /g | $m_f$ /g | $\mu_{rel}$ /wt.% |
| --- | --- | --- | --- |
| s1 | 0.01727 | 0.00754 | 44 |
| s2 | 0.01516 | 0.00604 | 40 |
| s3 | 0.0174 | 0.0032 | 18 |
| | | Average ± S.D. | 34 ± 14 |
| 48 h | $m_d$ /g | $m_f$ /g | $\mu_{rel}$ /wt.% |

| | $m_d$ /g | $m_f$ /g | $\mu_{rel}$ /wt.% |
|---|---|---|---|
| **s1** | 0.01583 | 0.00104 | 7 |
| **s2** | 0.01248 | 0.00075 | 6 |
| **s3** | 0.0211 | 0.00406 | 19 |
| | | **Average ± S.D.** | 11 ± 7 |
| **72 h** | $m_d$ /g | $m_f$ /g | $\mu_{rel}$ /wt.% |
| **s1** | 0.01347 | 0.00137 | 10 |
| **s2** | 0.01527 | 0.00238 | 16 |
| **s3** | 0.01553 | 0.00107 | 7 |
| | | **Average ± S.D.** | 11 ± 4 |

**Rat tail collagen-based fibre degradation**

| 24 h | $m_d$ /g | $m_f$ /g | $\mu_{rel}$ /wt.% |
|---|---|---|---|
| **s1** | 0.01257 | 0.00122 | 10 |
| **s2** | 0.01163 | 0.00137 | 12 |
| **s3** | 0.01081 | 0.00082 | 8 |
| | | **Average ± S.D.** | 10 ± 2 |
| **48 h** | $m_d$ /g | $m_f$ /g | $\mu_{rel}$ /wt.% |
| **s1** | 0.01321 | 0.00104 | 8 |
| **s2** | 0.01877 | 0.00185 | 10 |
| **s3** | 0.01409 | 0.00054 | 4 |
| | | **Average ± S.D.** | 7 ± 3 |
| **72 h** | $m_d$ /g | $m_f$ /g | $\mu_{rel}$ /wt.% |
| **s1** | 0.01489 | 0.00075 | 5 |
| **s2** | 0.01383 | 0.00027 | 2 |
| **s3** | 0.01359 | 0.00002 | 0 |
| | | **Average ± S.D.** | 2 ± 2 |

**Rat tail collagen-based film degradation**

| 24 h | $m_d$ /g | $m_f$ /g | $\mu_{rel}$ /wt.% |
|---|---|---|---|
| **s1** | 0.00803 | 0.00089 | 11 |
| **s2** | 0.01196 | 0.00218 | 18 |
| **s3** | 0.00943 | 0.00094 | 10 |
| | | **Average ± S.D.** | 13 ± 4 |
| **48 h** | $m_d$ /g | $m_f$ /g | $\mu_{rel}$ /wt.% |
| **s1** | 0.01095 | 0.00094 | 9 |
| **s2** | 0.00832 | 0.0003 | 4 |
| **s3** | 0.00811 | 0.0004 | 5 |
| | | **Average ± S.D.** | 6 ± 3 |

| 72 h | $m_d$ /g | $m_f$ /g | $\mu_{rel}$ /wt.% |
|---|---|---|---|
| s1 | 0.01213 | 0 | 5 |
| s2 | 0.01352 | 0.00069 | 2 |
| s3 | 0.00975 | 0.00081 | 0 |
| | | Average ± S.D. | 4 ± 4 |

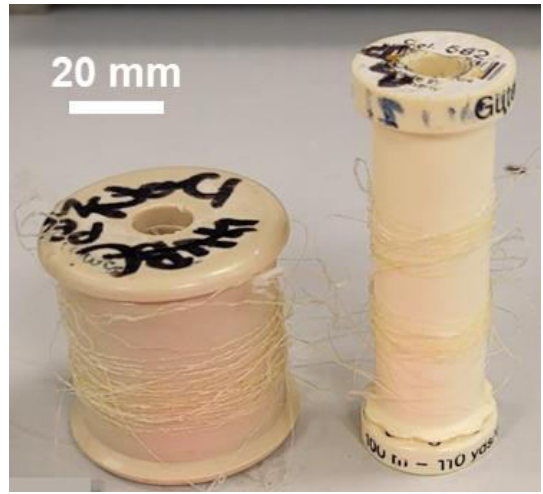

**Figure S3.** Spooled fibres exceeding 1 m in length collected after wet spinning of the I2959-supplemented collagen dope (0.5 wt.% I2959) in an I2959-free ethanol bath.

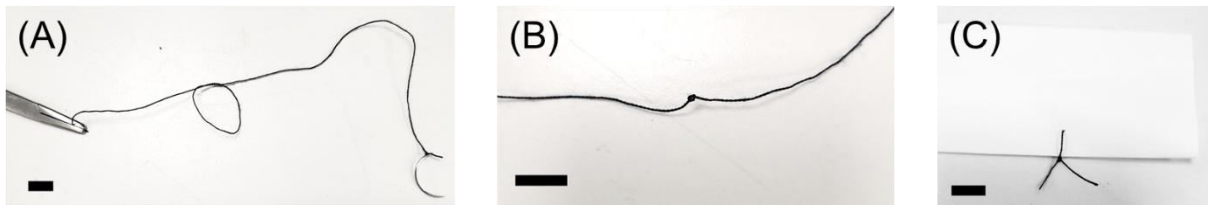

**Figure S4.** Photos of the commercial silk-braided suture control demonstrating respective knot-forming (A-B) and knot-tying (C) abilities. Scale bars ≈ 1 cm.

*Normality Test (10/06/2025 19:14:04)*

**Input Data**

| | Data | Range |
|---|---|---|
| Data | [Book1]Sheet1!Q | [1*:17*] |

There exist missing values in the input data

**Descriptive Statistics**

| | N Analysis | N Missing | Mean | Standard Deviation | SE of Mean |
|---|---|---|---|---|---|
| Q | 3 | 14 | 19.1527 | 3.11257 | 1.79704 |

**NormalityTest**

*Shapiro-Wilk*

| | DF | Statistic | p-value | Decision at level(5%) |
|---|---|---|---|---|
| Q | 3 | 0.82636 | 0.17911 | Can't reject normality |

Q: At the 0.05 level, the data was significantly drawn from a normally distributed population.

**Figure S5.** Normality test run on Origin software for the swelling ratio (SR) data obtained with the UV-cured thin films.

*Normality Test (10/06/2025 19:12:40*

**Input Data**

| | Data | Range |
|---|---|---|
| Data | [Book1]Sheet1!O"unloded" | [1*:6*] |

There exist missing values in the input data.

**Descriptive Statistics**

| | N Analysis | N Missing | Mean | Standard Deviation | SE of Mean |
|---|---|---|---|---|---|
| unloded | 5 | 1 | 4.91752 | 0.73221 | 0.32746 |

**NormalityTest**

*Shapiro-Wilk*

| | DF | Statistic | p-value | Decision at level(5%) |
|---|---|---|---|---|
| unloded | 5 | 0.98781 | 0.97146 | Can't reject normality |

unloded: At the 0.05 level, the data was significantly drawn from a normally distributed population.

**Figure S6.** Normality test run on Origin software for the swelling ratio (SR) data obtained with the BTB-loaded films.

```
Normality Test (10/06/2025 19:12:27
Input Data
        Data                    Range
Data    [Book1]Sheet1!N"loaded"  [1*:6*]
Descriptive Statistics
         N Analysis  N Missing  Mean    Standard Deviation  SE of Mean
loaded        6          0      3.22821        0.78543         0.32065
NormalityTest
Shapiro-Wilk
         DF   Statistic  p-value  Decision at level(5%)
loaded   6    0.937      0.63512  Can't reject normality
```
loaded: At the 0.05 level, the data was significantly drawn from a normally distributed population.

**Figure S7.** Normality test run on Origin software for the swelling ratio (SR) data obtained with the BTB-loaded wet spun fibres.

```
Normality Test (10/06/2025 19:14:31
Input Data
        Data                Range
Data    [Book1]Sheet1!T     [1*:4*]
Descriptive Statistics
     N Analysis  N Missing  Mean   Standard Deviation  SE of Mean
T         4          0      20.25        1.89297          0.94648
NormalityTest
Shapiro-Wilk
     DF   Statistic  p-value  Decision at level(5%)
T    4    0.79065    0.08649  Can't reject normality
```
T: At the 0.05 level, the data was significantly drawn from a normally distributed population.

**Figure S8.** Normality test run on Origin software for denaturation temperature data obtained via Thermogravimetric analysis (TGA) of BTB-loaded fibres. The same test on the BTB-free samples was rejected given that only one sample had remaining mass.

```
Normality Test (10/06/2025 19:07:14)
Notes
Description    Perform Normality Test
User Name      sdjg
Operation Time 10/06/2025 19:07:14
Report Status  New Analysis Report
Data Filter    No

Input Data
       Data                                        Range
Data   [Book1]Sheet1!E"Gel content percent age"   [1*:3*]

Descriptive Statistics
                       N Analysis  N Missing  Mean     Standard Deviation  SE of Mean
Gel content percentage     3          0       0.87197        0.04665          0.02694

NormalityTest
Shapiro-Wilk
                       DF  Statistic  p-value  Decision at level(5%)
Gel content percentage  3   0.99956   0.95987  Can't reject normality

Gel content percentage: At the 0.05 level, the data was significantly drawn from a normally distributed population.
```

**Figure S9.** Normality test run on Origin software for the gel content (SR) data obtained with the thin films.

```
Normality Test (10/06/2025 19:10:10)
Input Data
       Data                                        Range
Data   [Book1]Sheet1!G"Gel content percent age"   [1*:3*]

Descriptive Statistics
                       N Analysis  N Missing  Mean     Standard Deviation  SE of Mean
Gel content percentage     3          0       0.85797        0.05865          0.03386

NormalityTest
Shapiro-Wilk
                       DF  Statistic  p-value  Decision at level(5%)
Gel content percentage  3   0.97725   0.71085  Can't reject normality

Gel content percentage: At the 0.05 level, the data was significantly drawn from a normally distributed population.
```

**Figure S10.** Normality test run on Origin software for the gel content (G) data obtained with the wet spun fibres.

```
Normality Test (04/09/2025 10:54:45)
Input Data
    Data                              Range
Data [Book1]Sheet1!EG"fibres 24"  [1*:3*]
     [Book1]Sheet1!EH"films 24"   [1*:3*]
```

**Descriptive Statistics**

|          | N Analysis | N Missing | Mean     | Standard Deviation | SE of Mean |
|----------|------------|-----------|----------|--------------------|------------|
| fibres 24| 3          | 0         | 9.69037  | 2.0972             | 1.21082    |
| films 24 | 3          | 0         | 13.09302 | 4.48136            | 2.58731    |

**Normality Test**
**Shapiro-Wilk**

|          | DF | Statistic | p-value | Decision at level(5%)  |
|----------|----|-----------|---------|------------------------|
| fibres 24| 3  | 0.99996   | 0.98795 | Can't reject normality |
| films 24 | 3  | 0.84918   | 0.23826 | Can't reject normality |

fibres 24: At the 0.05 level, the data was significantly drawn from a normally distributed population.
films 24: At the 0.05 level, the data was significantly drawn from a normally distributed population.

**Figure S11.** Normality test run on Origin software for the 24-hour data of relative mass ($\mu_{rel}$) obtained with the thin films and wet spun fibres.

```
Normality Test (04/09/2025 10:54:52)
Input Data
    Data                              Range
Data [Book1]Sheet1!EJ"fibres 48"  [1*:3*]
     [Book1]Sheet1!EK"films 48"   [1*:3*]
```

**Descriptive Statistics**

|          | N Analysis | N Missing | Mean    | Standard Deviation | SE of Mean |
|----------|------------|-----------|---------|--------------------|------------|
| fibres 48| 3          | 0         | 7.18716 | 3.0698             | 1.77235    |
| films 48 | 3          | 0         | 5.70748 | 2.57831            | 1.48859    |

**Normality Test**
**Shapiro-Wilk**

|          | DF | Statistic | p-value | Decision at level(5%)  |
|----------|----|-----------|---------|------------------------|
| fibres 48| 3  | 0.96258   | 0.62823 | Can't reject normality |
| films 48 | 3  | 0.93219   | 0.49685 | Can't reject normality |

fibres 48: At the 0.05 level, the data was significantly drawn from a normally distributed population.
films 48: At the 0.05 level, the data was significantly drawn from a normally distributed population.

**Figure S12.** Normality test run on Origin software for the 48-hour data of relative mass ($\mu_{rel}$) obtained with the thin films and wet spun fibres.

**Normality Test (04/09/2025 10:55:00)**

**Input Data**

| | Data | Range |
|---|---|---|
| Data | [Book1]Sheet1!EM"fibres 72" | [1*:3*] |
| | [Book1]Sheet1!EN"films 72" | [1*:3*] |

**Descriptive Statistics**

| | N Analysis | N Missing | Mean | Standard Deviation | SE of Mean |
|---|---|---|---|---|---|
| fibres 72 | 3 | 0 | 2.37879 | 2.47263 | 1.42757 |
| films 72 | 3 | 0 | 4.47041 | 4.18988 | 2.41903 |

**NormalityTest**
**Shapiro-Wilk**

| | DF | Statistic | p-value | Decision at level(5%) |
|---|---|---|---|---|
| fibres 72 | 3 | 0.97768 | 0.71362 | Can't reject normality |
| films 72 | 3 | 0.98287 | 0.74934 | Can't reject normality |

fibres 72: At the 0.05 level, the data was significantly drawn from a normally distributed population.
films 72: At the 0.05 level, the data was significantly drawn from a normally distributed population.

**Figure S13.** Normality test run on Origin software for the 72-hour data of relative mass ($\mu_{rel}$) obtained with the thin films and wet spun fibres.

**Normality Test (03/07/2025 00:01:48)**

**Input Data**

| | Data | Range |
|---|---|---|
| Data | [Book1]Sheet1!EB"films" | [1*:27*] |
| | [Book1]Sheet1!EC"fibres" | [1*:3*] |

There exist missing values in the input data.

**NormalityTest**
**Shapiro-Wilk**

| | DF | Statistic | p-value | Decision at level(5%) |
|---|---|---|---|---|
| films | 24 | 0.90698 | 0.03034 | Reject normality |
| fibres | 3 | 0.99994 | 0.9853 | Can't reject normality |

films: At the 0.05 level, the data was not significantly drawn from a normally distributed population.
fibres: At the 0.05 level, the data was significantly drawn from a normally distributed population.

**Figure S14.** Normality test run on Origin software for the BTB release measured with the fibres and films following 12-hour incubation in an acidic (pH 5) McIlvaine solution. Films normality rejected- varied release profiles).

*Normality Test (02/07/2025 23:55:13)*
**Input Data**

| | Data | Range |
|---|---|---|
| Data | [Book1]Sheet1!DX"films" | [1*:6*] |
| | [Book1]Sheet1!DY"fibres" | [1*:3*] |

*Descriptive Statistics*

| | N Analysis | N Missing | Mean | Standard Deviation | SE of Mean |
|---|---|---|---|---|---|
| films | 6 | 0 | 12.65184 | 3.03431 | 1.23875 |
| fibres | 3 | 0 | 12.45957 | 2.64093 | 1.52474 |

*NormalityTest*
*Shapiro-Wilk*

| | DF | Statistic | p-value | Decision at level(5%) |
|---|---|---|---|---|
| films | 6 | 0.96943 | 0.8886 | Can't reject normality |
| fibres | 3 | 0.96429 | 0.63689 | Can't reject normality |

films: At the 0.05 level, the data was significantly drawn from a normally distributed population.
fibres: At the 0.05 level, the data was significantly drawn from a normally distributed population.

**Figure S15.** Normality test run on Origin software for the BTB release measured with the fibres and films following 12-hour incubation in an alkaline (pH 8) McIlvaine solution.

*Normality Test (15/09/2025 11:59:45*
**Input Data**

| | Data | Range |
|---|---|---|
| Data | [Book1]Sheet1!EP"films 24" | [1*:27*] |
| | [Book1]Sheet1!EQ"fibresw 24" | [1*:3*] |

There exist missing values in the input data.

**NormalityTest**
**Shapiro-Wilk**

| | DF | Statistic | p-value | Decision at level(5%) |
|---|---|---|---|---|
| films 24 | 24 | 0.9616 | 0.47142 | Can't reject normality |
| fibresw 24 | 3 | 0.98095 | 0.73554 | Can't reject normality |

films 24: At the 0.05 level, the data was significantly drawn from a normally distributed population.
fibresw 24: At the 0.05 level, the data was significantly drawn from a normally distributed population.

**Figure S16.** Normality test run on Origin software for the BTB release measured with the fibres and films following 24-hour incubation in an alkaline (pH 8) McIlvaine solution.

**Normality Test (15/09/2025 12:02:54)**

**Input Data**

| | Data | Range |
|---|---|---|
| Data | [Book1]Sheet1!ES"films 24" | [1*:27*] |
| | [Book1]Sheet1!ET"fibres 24" | [1*:3*] |

There exist missing values in the input data.

**Descriptive Statistics**

| | N Analysis | N Missing | Mean | Standard Deviation | SE of Mean |
|---|---|---|---|---|---|
| films 24 | 24 | 3 | 9.06805 | 5.52665 | 1.12812 |
| fibres 24 | 3 | 0 | 11.99853 | 2.15939 | 1.24672 |

**NormalityTest**
**Shapiro-Wilk**

| | DF | Statistic | p-value | Decision at level(5%) |
|---|---|---|---|---|
| films 24 | 24 | 0.90535 | 0.02799 | Reject normality |
| fibres 24 | 3 | 0.92308 | 0.46326 | Can't reject normality |

films 24: At the 0.05 level, the data was not significantly drawn from a normally distribu
fibres 24: At the 0.05 level, the data was significantly drawn from a normally distributed

**Figure S17.** Normality test run on Origin software for the BTB release measured with the fibres and films following 24-hour incubation in an acidic (pH 5) McIlvaine solution. Films normality rejected- varied release profiles).

## ANOVAOneWay (10/06/2025 19:31:00)

### Input Data

|  | Data | Range |
|---|---|---|
| loaded | [Book1]Sheet1!N"loaded" | [1*:6*] |
| sr thin films | [Book1]Sheet1!R"sr thin films" | [1*:7*] |

### Descriptive Statistics

|  | N Analysis | N Missing | Mean | Standard Deviation | SE of Mean |
|---|---|---|---|---|---|
| loaded | 6 | 0 | 3.22821 | 0.78543 | 0.32065 |
| sr thin films | 7 | 0 | 21.34965 | 4.02673 | 1.52196 |

### ANOVA

#### Overall ANOVA

|  | DF | Sum of Squares | Mean Square | F Value | Prob>F |
|---|---|---|---|---|---|
| Model | 1 | 1060.94116 | 1060.94116 | 116.27121 | <0.0001 |
| Error | 11 | 100.37182 | 9.12471 |  |  |
| Total | 12 | 1161.31297 |  |  |  |

Null Hypothesis: The means of all levels are equal.
Alternative Hypothesis: The means of one or more levels are different.
**At the 0.05 level, the population means are significantly different.**

### Fit Statistics

| R-Square | Coeff Var | Root MSE | Data Mean |
|---|---|---|---|
| 0.91357 | 0.23261 | 3.02071 | 12.98591 |

### Means Comparisons

#### Bonferroni Test

|  | MeanDiff | SEM | t Value | Prob | Alpha | Sig | LCL | UCL |
|---|---|---|---|---|---|---|---|---|
| sr thin films loaded | 18.12144 | 1.68057 | 10.78291 | <0.0001 | 0.05 | 1 | 14.42253 | 21.82035 |

#### Tukey Test

|  | MeanDiff | SEM | q Value | Prob | Alpha | Sig | LCL | UCL |
|---|---|---|---|---|---|---|---|---|
| sr thin films loaded | 18.12144 | 1.68057 | 15.24934 | <0.0001 | 0.05 | 1 | 14.42253 | 21.82035 |

#### Bonholm Test

|  | MeanDiff | SEM | t Value | Prob | Alpha | Sig |
|---|---|---|---|---|---|---|
| sr thin films loaded | 18.12144 | 1.68057 | 10.78291 | <0.0001 | 0.05 | 1 |

### Grouping Letters Table

#### Bonferroni Test

|  | Mean | Groups |
|---|---|---|
| sr thin films | 21.34965 | A |
| loaded | 3.22821 | B |

Means that do not share a letter are significantly different

#### Tukey Test

|  | Mean | Groups |
|---|---|---|
| sr thin films | 21.34965 | A |
| loaded | 3.22821 | B |

Means that do not share a letter are significantly different

#### Bonholm Test

|  | Mean | Groups |
|---|---|---|
| sr thin films | 21.34965 | A |
| loaded | 3.22821 | B |

Means that do not share a letter are significantly different

Sig equals 1 indicates that the difference of the means is significant at the 0.05 level.
Sig equals 0 indicates that the difference of the means is not significant at the 0.05 level.

**Figure S18.** One-way ANOVA run on Origin software for the swelling ratio (SR) data obtained with BTB-loaded fibres and BTB-loaded films.

## ANOVAOneWay (10/06/2025 19:31:57)

### Input Data

|  | Data | Range |
|---|---|---|
| unloded | [Book1]Sheet1!O"unloded" | [1*:6*] |
| sr thin films | [Book1]Sheet1!R"sr thin films" | [1*:7*] |

There exist missing values in the input data.

### Descriptive Statistics

|  | N Analysis | N Missing | Mean | Standard Deviation | SE of Mean |
|---|---|---|---|---|---|
| unloded | 5 | 1 | 4.91752 | 0.73221 | 0.32746 |
| sr thin films | 7 | 0 | 21.34965 | 4.02673 | 1.52196 |

### ANOVA

#### Overall ANOVA

|  | DF | Sum of Squares | Mean Square | F Value | Prob>F |
|---|---|---|---|---|---|
| Model | 1 | 787.54353 | 787.54353 | 79.20437 | <0.0001 |
| Error | 10 | 99.43183 | 9.94318 |  |  |
| Total | 11 | 886.97535 |  |  |  |

Null Hypothesis: The means of all levels are equal.
Alternative Hypothesis: The means of one or more levels are different.
**At the 0.05 level, the population means are significantly different.**

### Fit Statistics

| R-Square | Coeff Var | Root MSE | Data Mean |
|---|---|---|---|
| 0.8879 | 0.21742 | 3.15328 | 14.50293 |

### Means Comparisons

#### Bonferroni Test

|  | MeanDiff | SEM | t Value | Prob | Alpha | Sig | LCL | UCL |
|---|---|---|---|---|---|---|---|---|
| sr thin films  unloded | 16.43213 | 1.84637 | 8.89968 | <0.0001 | 0.05 | 1 | 12.31816 | 20.54611 |

#### Tukey Test

|  | MeanDiff | SEM | q Value | Prob | Alpha | Sig | LCL | UCL |
|---|---|---|---|---|---|---|---|---|
| sr thin films  unloded | 16.43213 | 1.84637 | 12.58605 | <0.0001 | 0.05 | 1 | 12.31815 | 20.54611 |

#### Bonholm Test

|  | MeanDiff | SEM | t Value | Prob | Alpha | Sig |
|---|---|---|---|---|---|---|
| sr thin films  unloded | 16.43213 | 1.84637 | 8.89968 | <0.0001 | 0.05 | 1 |

### Grouping Letters Table

#### Bonferroni Test

|  | Mean | Groups |
|---|---|---|
| sr thin films | 21.34965 | A |
| unloded | 4.91752 | B |

Means that do not share a letter are significantly different

#### Tukey Test

|  | Mean | Groups |
|---|---|---|
| sr thin films | 21.34965 | A |
| unloded | 4.91752 | B |

Means that do not share a letter are significantly different

#### Bonholm Test

|  | Mean | Groups |
|---|---|---|
| sr thin films | 21.34965 | A |
| unloded | 4.91752 | B |

Means that do not share a letter are significantly different

Sig equals 1 indicates that the difference of the means is significant at the 0.05 level.
Sig equals 0 indicates that the difference of the means is not significant at the 0.05 level.

**Figure S19.** One-way ANOVA run on Origin software for the swelling ratio (SR) measured with BTB-free fibres and BTB-free films.

ANOVAOneWay (10/06/2025 19:30:12)

### Input Data

|  | Data | Range |
|---|---|---|
| loaded | [Book1]Sheet1!N"loaded" | [1*:6*] |
| unloded | [Book1]Sheet1!O"unloded" | [1*:6*] |

There exist missing values in the input data.

### Descriptive Statistics

|  | N Analysis | N Missing | Mean | Standard Deviation | SE of Mean |
|---|---|---|---|---|---|
| loaded | 6 | 0 | 3.22821 | 0.78543 | 0.32065 |
| unloded | 5 | 1 | 4.91752 | 0.73221 | 0.32746 |

### ANOVA

#### Overall ANOVA

|  | DF | Sum of Squares | Mean Square | F Value | Prob>F |
|---|---|---|---|---|---|
| Model | 1 | 7.78299 | 7.78299 | 13.39562 | 0.00524 |
| Error | 9 | 5.22909 | 0.58101 |  |  |
| Total | 10 | 13.01207 |  |  |  |

Null Hypothesis: The means of all levels are equal.
Alternative Hypothesis: The means of one or more levels are different.
**At the 0.05 level, the population means are significantly different.**

#### Fit Statistics

| R-Square | Coeff Var | Root MSE | Data Mean |
|---|---|---|---|
| 0.59814 | 0.19075 | 0.76224 | 3.99608 |

### Means Comparisons

#### Bonferroni Test

|  | MeanDiff | SEM | t Value | Prob | Alpha | Sig | LCL | UCL |
|---|---|---|---|---|---|---|---|---|
| unloded loaded | 1.68931 | 0.46156 | 3.66 | 0.00524 | 0.05 | 1 | 0.64519 | 2.73343 |

#### Tukey Test

|  | MeanDiff | SEM | q Value | Prob | Alpha | Sig | LCL | UCL |
|---|---|---|---|---|---|---|---|---|
| unloded loaded | 1.68931 | 0.46156 | 5.17603 | 0.00524 | 0.05 | 1 | 0.64519 | 2.73343 |

#### Bonholm Test

|  | MeanDiff | SEM | t Value | Prob | Alpha | Sig |
|---|---|---|---|---|---|---|
| unloded loaded | 1.68931 | 0.46156 | 3.66 | 0.00524 | 0.05 | 1 |

### Grouping Letters Table

#### Bonferroni Test

|  | Mean | Groups |
|---|---|---|
| unloded | 4.91752 | A |
| loaded | 3.22821 | B |

Means that do not share a letter are significantly different

#### Tukey Test

|  | Mean | Groups |
|---|---|---|
| unloded | 4.91752 | A |
| loaded | 3.22821 | B |

Means that do not share a letter are significantly different

#### Bonholm Test

|  | Mean | Groups |
|---|---|---|
| unloded | 4.91752 | A |
| loaded | 3.22821 | B |

Means that do not share a letter are significantly different

Sig equals 1 indicates that the difference of the means is significant at the 0.05 level.
Sig equals 0 indicates that the difference of the means is not significant at the 0.05 level.

**Figure S20.** One-way ANOVA run on Origin software for the swelling ratio (SR) data measured with BTB-loaded and BTB-free fibres.

ANOVAOneWay (15/09/2025 11:47:33)

**Input Data**

| | Data | Range |
|---|---|---|
| TGA load | [Book1]Sheet1!T"TGA load" | [1*:4*] |
| TGA unload | [Book1]Sheet1!U"TGA unload" | [1*:1*] |

**Descriptive Statistics**

| | N Analysis | N Missing | Mean | Standard Deviation | SE of Mean |
|---|---|---|---|---|---|
| TGA load | 4 | 0 | 20.25 | 1.89297 | 0.94648 |
| TGA unload | 1 | 0 | 0.61 | -- | -- |

## ANOVA
### Overall ANOVA

| | DF | Sum of Squares | Mean Square | F Value | Prob>F |
|---|---|---|---|---|---|
| Model | 1 | 308.58368 | 308.58368 | 86.11638 | 0.00265 |
| Error | 3 | 10.75 | 3.58333 | | |
| Total | 4 | 319.33368 | | | |

Null Hypothesis: The means of all levels are equal.
Alternative Hypothesis: The means of one or more levels are different.
At the 0.05 level, the population means are significantly different.

**Fit Statistics**

| R-Square | Coeff Var | Root MSE | Data Mean |
|---|---|---|---|
| 0.96634 | 0.11598 | 1.89297 | 16.322 |

### Means Comparisons
#### Bonferroni Test

| | MeanDiff | SEM | t Value | Prob | Alpha | Sig | LCL | UCL |
|---|---|---|---|---|---|---|---|---|
| TGA unload TGA load | -19.64 | 2.1164 | -9.27989 | 0.00265 | 0.05 | 1 | -26.37535 | -12.90465 |

#### Tukey Test

| | MeanDiff | SEM | q Value | Prob | Alpha | Sig | LCL | UCL |
|---|---|---|---|---|---|---|---|---|
| TGA unload TGA load | -19.64 | 2.1164 | 13.12375 | 0.00265 | 0.05 | 1 | -26.37535 | -12.90465 |

#### Bonholm Test

| | MeanDiff | SEM | t Value | Prob | Alpha | Sig |
|---|---|---|---|---|---|---|
| TGA unload TGA load | -19.64 | 2.1164 | -9.27989 | 0.00265 | 0.05 | 1 |

### Grouping Letters Table
#### Bonferroni Test

| | Mean | Groups |
|---|---|---|
| TGA load | 20.25 | A |
| TGA unload | 0.61 | B |

Means that do not share a letter are significantly different

#### Tukey Test

| | Mean | Groups |
|---|---|---|
| TGA load | 20.25 | A |
| TGA unload | 0.61 | B |

Means that do not share a letter are significantly different

#### Bonholm Test

| | Mean | Groups |
|---|---|---|
| TGA load | 20.25 | A |
| TGA unload | 0.61 | B |

Means that do not share a letter are significantly different

Sig equals 1 indicates that the difference of the means is significant at the 0.05 level.
Sig equals 0 indicates that the difference of the means is not significant at the 0.05 level.

**Figure S21.** One-way ANOVA run on Origin software for the denaturation temperature ($T_d$) data obtained via thermogravimetric analysis (TGA) of BTB-loaded and BTB-free fibres.

ANOVAOneWay (10/06/2025 19:15:46)

Input Data

|  | Data | Range |
|---|---|---|
| Gel content percentage | [Book1]Sheet1!E"Gel content percentage" | [1*:3*] |
| Gel content percentage | [Book1]Sheet1!G"Gel content percentage" | [1*:3*] |

Descriptive Statistics

|  | N Analysis | N Missing | Mean | Standard Deviation | SE of Mean |
|---|---|---|---|---|---|
| Gel content percentage | 3 | 0 | 0.87197 | 0.04665 | 0.02694 |
|  | 3 | 0 | 0.85797 | 0.05865 | 0.03386 |

ANOVA

Overall ANOVA

|  | DF | Sum of Squares | Mean Square | F Value | Prob>F |
|---|---|---|---|---|---|
| Model | 1 | 2.94075E-4 | 2.94075E-4 | 0.10472 | 0.76245 |
| Error | 4 | 0.01123 | 0.00281 |  |  |
| Total | 5 | 0.01153 |  |  |  |

Null Hypothesis: The means of all levels are equal.
Alternative Hypothesis: The means of one or more levels are different.
At the 0.05 level, the population means are not significantly different.

Fit Statistics

| R-Square | Coeff Var | Root MSE | Data Mean |
|---|---|---|---|
| 0.02551 | 0.06127 | 0.05299 | 0.86497 |

Means Comparisons

Bonferroni Test

|  | MeanDiff | SEM | t Value | Prob | Alpha | Sig | LCL | UCL |
|---|---|---|---|---|---|---|---|---|
| Gel content percentage Gel content percentage | -0.014 | 0.04327 | -0.3236 | 0.76245 | 0.05 | 0 | -0.13414 | 0.10613 |

Tukey Test

|  | MeanDiff | SEM | q Value | Prob | Alpha | Sig | LCL | UCL |
|---|---|---|---|---|---|---|---|---|
| Gel content percentage Gel content percentage | -0.014 | 0.04327 | 0.45764 | 0.76246 | 0.05 | 0 | -0.13413 | 0.10613 |

Bonholm Test

|  | MeanDiff | SEM | t Value | Prob | Alpha | Sig |
|---|---|---|---|---|---|---|
| Gel content percentage Gel content percentage | -0.014 | 0.04327 | -0.3236 | 0.76245 | 0.05 | 0 |

Grouping Letters Table

Bonferroni Test

|  | Mean | Groups |
|---|---|---|
| Gel content percentage | 0.87197 | A |
|  | 0.85797 | A |

Means that do not share a letter are significantly different

Tukey Test

|  | Mean | Groups |
|---|---|---|
| Gel content percentage | 0.87197 | A |
|  | 0.85797 | A |

Means that do not share a letter are significantly different

Bonholm Test

|  | Mean | Groups |
|---|---|---|
| Gel content percentage | 0.87197 | A |
|  | 0.85797 | A |

Means that do not share a letter are significantly different

Sig equals 1 indicates that the difference of the means is significant at the 0.05 level.
Sig equals 0 indicates that the difference of the means is not significant at the 0.05 level.

**Figure S22.** One-way ANOVA run on Origin software for the gel content (G) data obtained with the thin films and wet spun fibres.

**ANOVAOneWay (04/09/2025 10:46:01)**

**Input Data**

| | Data | Range |
|---|---|---|
| 616nm | [Book1]Sheet1!DX"films" | [1*:6*] |
| DY | [Book1]Sheet1!DY"fibres" | [1*:3*] |

**Descriptive Statistics**

| | N Analysis | N Missing | Mean | Standard Deviation | SE of Mean |
|---|---|---|---|---|---|
| films | 6 | 0 | 12.65184 | 3.03431 | 1.23875 |
| fibres | 3 | 0 | 12.45957 | 2.64093 | 1.52474 |

**ANOVA**

**Overall ANOVA**

| | DF | Sum of Squares | Mean Square | F Value | Prob>F |
|---|---|---|---|---|---|
| Model | 1 | 0.07393 | 0.07393 | 0.00863 | 0.9286 |
| Error | 7 | 59.98425 | 8.56918 | | |
| Total | 8 | 60.05819 | | | |

Null Hypothesis: The means of all levels are equal.
Alternative Hypothesis: The means of one or more levels are different.
At the 0.05 level, the population means are not significantly different.

**Fit Statistics**

| R-Square | Coeff Var | Root MSE | Data Mean |
|---|---|---|---|
| 0.00123 | 0.23255 | 2.92732 | 12.58775 |

**Means Comparisons**

**Bonferroni Test**

| | MeanDiff | SEM | t Value | Prob | Alpha | Sig | LCL | UCL |
|---|---|---|---|---|---|---|---|---|
| DY 616nm | -0.19227 | 2.06993 | -0.09289 | 0.9286 | 0.05 | 0 | -5.08686 | 4.70233 |

**Tukey Test**

| | MeanDiff | SEM | q Value | Prob | Alpha | Sig | LCL | UCL |
|---|---|---|---|---|---|---|---|---|
| DY 616nm | -0.19227 | 2.06993 | 0.13136 | 0.9286 | 0.05 | 0 | -5.08686 | 4.70233 |

**Bonholm Test**

| | MeanDiff | SEM | t Value | Prob | Alpha | Sig |
|---|---|---|---|---|---|---|
| DY 616nm | -0.19227 | 2.06993 | -0.09289 | 0.9286 | 0.05 | 0 |

**Grouping Letters Table**

**Bonferroni Test**

| | Mean | Groups |
|---|---|---|
| 616nm | 12.65184 | A |
| DY | 12.45957 | A |

Means that do not share a letter are significantly different

**Tukey Test**

| | Mean | Groups |
|---|---|---|
| 616nm | 12.65184 | A |
| DY | 12.45957 | A |

Means that do not share a letter are significantly different

**Bonholm Test**

| | Mean | Groups |
|---|---|---|
| 616nm | 12.65184 | A |
| DY | 12.45957 | A |

Means that do not share a letter are significantly different

Sig equals 1 indicates that the difference of the means is significant at the 0.05 level.
Sig equals 0 indicates that the difference of the means is not significant at the 0.05 level.

**Figure S23.** One-way ANOVA run on Origin software for the 12-hour BTB release measured following incubation of either wet spun fibres or thin films in an alkaline (pH 8) McIlvaine solution.

## ANOVAOneWay (04/09/2025 11:22:15)

### Input Data

| | Data | Range |
|---|---|---|
| 432nm | [Book1]Sheet1!EB"films" | [1*:27*] |
| EC | [Book1]Sheet1!EC"fibres" | [1*:3*] |

There exist missing values in the input data.

### Descriptive Statistics

| | N Analysis | N Missing | Mean | Standard Deviation | SE of Mean |
|---|---|---|---|---|---|
| films | 24 | 3 | 8.71507 | 5.26341 | 1.07439 |
| fibres | 3 | 0 | 36.6612 | 7.28068 | 4.2035 |

### ANOVA

#### Overall ANOVA

| | DF | Sum of Squares | Mean Square | F Value | Prob>F |
|---|---|---|---|---|---|
| Model | 1 | 2082.62978 | 2082.62978 | 70.05644 | <0.0001 |
| Error | 25 | 743.19716 | 29.72789 | | |
| Total | 26 | 2825.82693 | | | |

Null Hypothesis: The means of all levels are equal.
Alternative Hypothesis: The means of one or more levels are different.
At the 0.05 level, the population means are significantly different.

#### Fit Statistics

| R-Square | Coeff Var | Root MSE | Data Mean |
|---|---|---|---|
| 0.737 | 0.46127 | 5.45233 | 11.8202 |

### Means Comparisons

#### Bonferroni Test

| | MeanDiff | SEM | t Value | Prob | Alpha | Sig | LCL | UCL |
|---|---|---|---|---|---|---|---|---|
| EC 432nm | 27.94613 | 3.33886 | 8.36997 | <0.0001 | 0.05 | 1 | 21.06963 | 34.82263 |

#### Tukey Test

| | MeanDiff | SEM | q Value | Prob | Alpha | Sig | LCL | UCL |
|---|---|---|---|---|---|---|---|---|
| EC 432nm | 27.94613 | 3.33886 | 11.83693 | <0.0001 | 0.05 | 1 | 21.0696 | 34.82266 |

#### Bonholm Test

| | MeanDiff | SEM | t Value | Prob | Alpha | Sig |
|---|---|---|---|---|---|---|
| EC 432nm | 27.94613 | 3.33886 | 8.36997 | <0.0001 | 0.05 | 1 |

### Grouping Letters Table

#### Bonferroni Test

| | Mean | Groups |
|---|---|---|
| EC | 36.6612 | A |
| 432nm | 8.71507 | B |

Means that do not share a letter are significantly different.

#### Tukey Test

| | Mean | Groups |
|---|---|---|
| EC | 36.6612 | A |
| 432nm | 8.71507 | B |

Means that do not share a letter are significantly different.

#### Bonholm Test

| | Mean | Groups |
|---|---|---|
| EC | 36.6612 | A |
| 432nm | 8.71507 | B |

Means that do not share a letter are significantly different.

Sig equals 1 indicates that the difference of the means is significant at the 0.05 level.
Sig equals 0 indicates that the difference of the means is not significant at the 0.05 level.

**Figure S24.** One-way ANOVA run on Origin software for the 12-hour BTB release measured following incubation of either wet spun fibres or thin films in an acidic (pH 5) McIlvain solution.

## ANOVAOneWay (15/09/2025 12:06:57)

### Input Data

| | Data | Range |
|---|---|---|
| films 24 | [Book1]Sheet1!EP"films 24" | [1*:27*] |
| fibres 24 | [Book1]Sheet1!EQ"fibres 24" | [1*:3*] |

There exist missing values in the input data.

### Descriptive Statistics

| | N Analysis | N Missing | Mean | Standard Deviation | SE of Mean |
|---|---|---|---|---|---|
| films 24 | 24 | 3 | 18.47621 | 5.89393 | 1.20309 |
| fibres 24 | 3 | 0 | 39.6381 | 5.68464 | 3.28203 |

### ANOVA

#### Overall ANOVA

| | DF | Sum of Squares | Mean Square | F Value | Prob>F |
|---|---|---|---|---|---|
| Model | 1 | 1194.20197 | 1194.20197 | 34.56995 | <0.0001 |
| Error | 25 | 863.61277 | 34.54451 | | |
| Total | 26 | 2057.81474 | | | |

Null Hypothesis: The means of all levels are equal.
Alternative Hypothesis: The means of one or more levels are different.
At the 0.05 level, the population means are significantly different.

### Fit Statistics

| R-Square | Coeff Var | Root MSE | Data Mean |
|---|---|---|---|
| 0.58033 | 0.2822 | 5.87746 | 20.82753 |

### Means Comparisons

#### Bonferroni Test

| | MeanDiff | SEM | t Value | Prob | Alpha | Sig | LCL | UCL |
|---|---|---|---|---|---|---|---|---|
| fibres 24  films 24 | 21.16189 | 3.59919 | 5.87962 | <0.0001 | 0.05 | 1 | 13.74922 | 28.57457 |

#### Tukey Test

| | MeanDiff | SEM | q Value | Prob | Alpha | Sig | LCL | UCL |
|---|---|---|---|---|---|---|---|---|
| fibres 24  films 24 | 21.16189 | 3.59919 | 8.31504 | <0.0001 | 0.05 | 1 | 13.74918 | 28.57461 |

#### Bonholm Test

| | MeanDiff | SEM | t Value | Prob | Alpha | Sig |
|---|---|---|---|---|---|---|
| fibres 24  films 24 | 21.16189 | 3.59919 | 5.87962 | <0.0001 | 0.05 | 1 |

### Grouping Letters Table

#### Bonferroni Test

| | Mean | Groups |
|---|---|---|
| fibres 24 | 39.6381 | A |
| films 24 | 18.47621 | B |

Means that do not share a letter are significantly different

#### Tukey Test

| | Mean | Groups |
|---|---|---|
| fibres 24 | 39.6381 | A |
| films 24 | 18.47621 | B |

Means that do not share a letter are significantly different

#### Bonholm Test

| | Mean | Groups |
|---|---|---|
| fibres 24 | 39.6381 | A |
| films 24 | 18.47621 | B |

Means that do not share a letter are significantly different

Sig equals 1 indicates that the difference of the means is significant at the 0.05 level.
Sig equals 0 indicates that the difference of the means is not significant at the 0.05 level.

**Figure S25.** One-way ANOVA run on Origin software for the 24-hour BTB release measured following incubation of either wet spun fibres or thin films in an alkaline (pH 8) McIlvaine solution.

## ANOVAOneWay (15/09/2025 12:09:48)
### Descriptive Statistics

|          | N Analysis | N Missing | Mean     | Standard Deviation | SE of Mean |
|----------|------------|-----------|----------|--------------------|------------|
| films 24 | 24         | 3         | 9.06805  | 5.52665            | 1.12812    |
| fibres 24| 3          | 0         | 11.99853 | 2.15939            | 1.24672    |

### ANOVA
#### Overall ANOVA

|       | DF | Sum of Squares | Mean Square | F Value | Prob>F  |
|-------|----|----------------|-------------|---------|---------|
| Model | 1  | 22.90052       | 22.90052    | 0.80428 | 0.37837 |
| Error | 25 | 711.83391      | 28.47336    |         |         |
| Total | 26 | 734.73443      |             |         |         |

Null Hypothesis: The means of all levels are equal.
Alternative Hypothesis: The means of one or more levels are different.
At the 0.05 level, the population means are not significantly different.

### Fit Statistics

| R-Square | Coeff Var | Root MSE | Data Mean |
|----------|-----------|----------|-----------|
| 0.03117  | 0.56805   | 5.33604  | 9.39366   |

### Means Comparisons
#### Bonferroni Test

|                    | MeanDiff | SEM     | t Value | Prob    | Alpha | Sig | LCL      | UCL     |
|--------------------|----------|---------|---------|---------|-------|-----|----------|---------|
| fibres 24  films 24| 2.93048  | 3.26765 | 0.89682 | 0.37837 | 0.05  | 0   | -3.79937 | 9.66032 |

#### Tukey Test

|                    | MeanDiff | SEM     | q Value | Prob    | Alpha | Sig | LCL     | UCL     |
|--------------------|----------|---------|---------|---------|-------|-----|---------|---------|
| fibres 24  films 24| 2.93048  | 3.26765 | 1.26829 | 0.37837 | 0.05  | 0   | -3.7994 | 9.66035 |

#### Bonholm Test

|                    | MeanDiff | SEM     | t Value | Prob    | Alpha | Sig |
|--------------------|----------|---------|---------|---------|-------|-----|
| fibres 24  films 24| 2.93048  | 3.26765 | 0.89682 | 0.37837 | 0.05  | 0   |

### Grouping Letters Table
#### Bonferroni Test

|           | Mean     | Groups |
|-----------|----------|--------|
| fibres 24 | 11.99853 | A      |
| films 24  | 9.06805  | A      |

Means that do not share a letter are significantly different

#### Tukey Test

|           | Mean     | Groups |
|-----------|----------|--------|
| fibres 24 | 11.99853 | A      |
| films 24  | 9.06805  | A      |

Means that do not share a letter are significantly different

#### Bonholm Test

|           | Mean     | Groups |
|-----------|----------|--------|
| fibres 24 | 11.99853 | A      |
| films 24  | 9.06805  | A      |

Means that do not share a letter are significantly different

Sig equals 1 indicates that the difference of the means is significant at the 0.05 level.
Sig equals 0 indicates that the difference of the means is not significant at the 0.05 level.

**Figure S26.** One-way ANOVA run on Origin software for the 24-hour BTB release measured following incubation of either wet spun fibres or thin films in an acidic (pH 5) McIlvain solution.

## ANOVAOneWay (04/09/2025 10:57:38)

### Input Data

| | Data | Range |
|---|---|---|
| fibres 24 | [Book1]Sheet1!EG"fibres 24" | [1*:3*] |
| films 24 | [Book1]Sheet1!EH"films 24" | [1*:3*] |

### Descriptive Statistics

| | N Analysis | N Missing | Mean | Standard Deviation | SE of Mean |
|---|---|---|---|---|---|
| fibres 24 | 3 | 0 | 9.69037 | 2.0972 | 1.21082 |
| films 24 | 3 | 0 | 13.09302 | 4.48136 | 2.58731 |

### ANOVA

#### Overall ANOVA

| | DF | Sum of Squares | Mean Square | F Value | Prob>F |
|---|---|---|---|---|---|
| Model | 1 | 17.36705 | 17.36705 | 1.41883 | 0.29944 |
| Error | 4 | 48.96159 | 12.2404 | | |
| Total | 5 | 66.32864 | | | |

Null Hypothesis: The means of all levels are equal.
Alternative Hypothesis: The means of one or more levels are different.
At the 0.05 level, the population means are not significantly different.

### Fit Statistics

| R-Square | Coeff Var | Root MSE | Data Mean |
|---|---|---|---|
| 0.26183 | 0.30712 | 3.49863 | 11.39169 |

### Means Comparisons

#### Bonferroni Test

| | MeanDiff | SEM | t Value | Prob | Alpha | Sig | LCL | UCL |
|---|---|---|---|---|---|---|---|---|
| films 24  fibres 24 | 3.40265 | 2.85662 | 1.19115 | 0.29944 | 0.05 | 0 | -4.52859 | 11.33389 |

#### Tukey Test

| | MeanDiff | SEM | q Value | Prob | Alpha | Sig | LCL | UCL |
|---|---|---|---|---|---|---|---|---|
| films 24  fibres 24 | 3.40265 | 2.85662 | 1.68454 | 0.29944 | 0.05 | 0 | -4.52851 | 11.33381 |

#### Bonholm Test

| | MeanDiff | SEM | t Value | Prob | Alpha | Sig |
|---|---|---|---|---|---|---|
| films 24  fibres 24 | 3.40265 | 2.85662 | 1.19115 | 0.29944 | 0.05 | 0 |

### Grouping Letters Table

#### Bonferroni Test

| | Mean | Groups |
|---|---|---|
| films 24 | 13.09302 | A |
| fibres 24 | 9.69037 | A |

Means that do not share a letter are significantly different

#### Tukey Test

| | Mean | Groups |
|---|---|---|
| films 24 | 13.09302 | A |
| fibres 24 | 9.69037 | A |

Means that do not share a letter are significantly different

#### Bonholm Test

| | Mean | Groups |
|---|---|---|
| films 24 | 13.09302 | A |
| fibres 24 | 9.69037 | A |

Means that do not share a letter are significantly different

Sig equals 1 indicates that the difference of the means is significant at the 0.05 level.
Sig equals 0 indicates that the difference of the means is not significant at the 0.05 level.

**Figure S27.** One-way ANOVA run on Origin software for the 24-hour data of relative mass ($\mu_{rel}$) obtained with the thin films and wet spun fibres.

## ANOVAOneWay (04/09/2025 10:57:44)

### Input Data

| | Data | Range |
|---|---|---|
| fibres 48 | [Book1]Sheet1!EJ"fibres 48" | [1*:3*] |
| films 48 | [Book1]Sheet1!EK"films 48" | [1*:3*] |

### Descriptive Statistics

| | N Analysis | N Missing | Mean | Standard Deviation | SE of Mean |
|---|---|---|---|---|---|
| fibres 48 | 3 | 0 | 7.18716 | 3.0698 | 1.77235 |
| films 48 | 3 | 0 | 5.70748 | 2.57831 | 1.48859 |

### ANOVA

#### Overall ANOVA

| | DF | Sum of Squares | Mean Square | F Value | Prob>F |
|---|---|---|---|---|---|
| Model | 1 | 3.2842 | 3.2842 | 0.4087 | 0.5574 |
| Error | 4 | 32.14274 | 8.03569 | | |
| Total | 5 | 35.42695 | | | |

Null Hypothesis: The means of all levels are equal.
Alternative Hypothesis: The means of one or more levels are different.
At the 0.05 level, the population means are not significantly different.

### Fit Statistics

| R-Square | Coeff Var | Root MSE | Data Mean |
|---|---|---|---|
| 0.0927 | 0.43968 | 2.83473 | 6.44732 |

### Means Comparisons

#### Bonferroni Test

| | MeanDiff | SEM | t Value | Prob | Alpha | Sig | LCL | UCL |
|---|---|---|---|---|---|---|---|---|
| films 48 fibres 48 | -1.47969 | 2.31455 | -0.6393 | 0.5574 | 0.05 | 0 | -7.9059 | 4.94652 |

#### Tukey Test

| | MeanDiff | SEM | q Value | Prob | Alpha | Sig | LCL | UCL |
|---|---|---|---|---|---|---|---|---|
| films 48 fibres 48 | -1.47969 | 2.31455 | 0.9041 | 0.5574 | 0.05 | 0 | -7.90583 | 4.94646 |

#### Bonholm Test

| | MeanDiff | SEM | t Value | Prob | Alpha | Sig |
|---|---|---|---|---|---|---|
| films 48 fibres 48 | -1.47969 | 2.31455 | -0.6393 | 0.5574 | 0.05 | 0 |

### Grouping Letters Table

#### Bonferroni Test

| | Mean | Groups |
|---|---|---|
| fibres 48 | 7.18716 | A |
| films 48 | 5.70748 | A |

Means that do not share a letter are significantly different

#### Tukey Test

| | Mean | Groups |
|---|---|---|
| fibres 48 | 7.18716 | A |
| films 48 | 5.70748 | A |

Means that do not share a letter are significantly different

#### Bonholm Test

| | Mean | Groups |
|---|---|---|
| fibres 48 | 7.18716 | A |
| films 48 | 5.70748 | A |

Means that do not share a letter are significantly different

Sig equals 1 indicates that the difference of the means is significant at the 0.05 level.
Sig equals 0 indicates that the difference of the means is not significant at the 0.05 level.

**Figure S28.** One-way ANOVA run on Origin software for the 48-hour data of relative mass ($\mu_{rel}$) obtained with the thin films and wet spun fibres.

## ANOVAOneWay (04/09/2025 10:57:50)

### Input Data

| | Data | Range |
|---|---|---|
| fibres 72 | [Book1]Sheet1!EM"fibres 72" | [1*:3*] |
| films 72 | [Book1]Sheet1!EN"films 72" | [1*:3*] |

### Descriptive Statistics

| | N Analysis | N Missing | Mean | Standard Deviation | SE of Mean |
|---|---|---|---|---|---|
| fibres 72 | 3 | 0 | 2.37879 | 2.47263 | 1.42757 |
| films 72 | 3 | 0 | 4.47041 | 4.18988 | 2.41903 |

### ANOVA

#### Overall ANOVA

| | DF | Sum of Squares | Mean Square | F Value | Prob>F |
|---|---|---|---|---|---|
| Model | 1 | 6.56231 | 6.56231 | 0.55451 | 0.49785 |
| Error | 4 | 47.33797 | 11.83449 | | |
| Total | 5 | 53.90028 | | | |

Null Hypothesis: The means of all levels are equal.
Alternative Hypothesis: The means of one or more levels are different.
At the 0.05 level, the population means are not significantly different.

#### Fit Statistics

| R-Square | Coeff Var | Root MSE | Data Mean |
|---|---|---|---|
| 0.12175 | 1.00453 | 3.44013 | 3.4246 |

### Means Comparisons

#### Bonferroni Test

| | MeanDiff | SEM | t Value | Prob | Alpha | Sig | LCL | UCL |
|---|---|---|---|---|---|---|---|---|
| films 72 fibres 72 | 2.09162 | 2.80885 | 0.74465 | 0.49785 | 0.05 | 0 | -5.70701 | 9.89025 |

#### Tukey Test

| | MeanDiff | SEM | q Value | Prob | Alpha | Sig | LCL | UCL |
|---|---|---|---|---|---|---|---|---|
| films 72 fibres 72 | 2.09162 | 2.80885 | 1.0531 | 0.49785 | 0.05 | 0 | -5.70693 | 9.89017 |

#### Bonholm Test

| | MeanDiff | SEM | t Value | Prob | Alpha | Sig |
|---|---|---|---|---|---|---|
| films 72 fibres 72 | 2.09162 | 2.80885 | 0.74465 | 0.49785 | 0.05 | 0 |

### Grouping Letters Table

#### Bonferroni Test

| | Mean | Groups |
|---|---|---|
| films 72 | 4.47041 | A |
| fibres 72 | 2.37879 | A |

Means that do not share a letter are significantly different

#### Tukey Test

| | Mean | Groups |
|---|---|---|
| films 72 | 4.47041 | A |
| fibres 72 | 2.37879 | A |

Means that do not share a letter are significantly different

#### Bonholm Test

| | Mean | Groups |
|---|---|---|
| films 72 | 4.47041 | A |
| fibres 72 | 2.37879 | A |

Means that do not share a letter are significantly different

Sig equals 1 indicates that the difference of the means is significant at the 0.05 level.
Sig equals 0 indicates that the difference of the means is not significant at the 0.05 level.

**Figure S29.** One-way ANOVA run on Origin software for the 72-hour data of relative mass ($\mu_{rel}$) obtained with the thin films and wet spun fibres.